\documentclass[12pt]{article}
\usepackage{amssymb, amsmath, hyperref,verbatim}
\usepackage{graphicx}
\usepackage[latin1]{inputenc}
\usepackage{verbatim}
\def\ltwid{\mathrel{\raise.3ex\hbox{$<$\kern-.75em\lower1ex\hbox{$\sim$}}}}
\def \be{\begin{equation}}
\def \ee{\end{equation}}
\def \bea{\begin{eqnarray}}
\def \eea{\end{eqnarray}}

\def \f{\frac}

\def\ltwid{\mathrel{\raise.3ex\hbox{$<$\kern-.75em\lower1ex\hbox{$\sim$}}}}

\def\square{\kern1pt\vbox{\hrule height 1.2pt\hbox{\vrule width 1.2pt\hskip 3pt
   \vbox{\vskip 6pt}\hskip 3pt\vrule width 0.6pt}\hrule height 0.6pt}\kern1pt}

\begin{document}
\begin{titlepage}
\begin{flushright}
ITP-UU-08/38, SPIN-08/29
\end{flushright}

\vspace{0.5cm}

\begin{center}
\bf{The graviton one-loop effective action in cosmological
space-times with constant deceleration}
\end{center}

\vspace{0.3cm} \vspace{0.1cm}

\begin{center}
T. M. Janssen$^{*}$, T. Prokopec$^{\ddagger}$
\end{center}
\begin{center}
\it{Institute for Theoretical Physics \& Spinoza Institute,
Utrecht University,\\
Leuvenlaan 4, Postbus 80.195, 3508 TD Utrecht, THE NETHERLANDS}
\end{center}

\vspace{0.3cm}

\begin{center}
ABSTRACT
\end{center}
We consider the quantum Friedmann equations which include one-loop
vacuum fluctuations due to gravitons and scalar field matter in a
FLRW background with constant $\epsilon=-{\dot{H}}/{H^2}$. After
several field redefinitions, to remove the mixing between the
gravitational and matter degrees of freedom, we can construct the
one loop correction to the Friedmann equations. Due to
cosmological particle creation, the propagators needed in such a
calculation are typically infrared divergent. In this paper we
construct the graviton and matter propagators, making use of the
recent construction of the infrared finite scalar propagators
calculated on a compact spatial manifold in~\cite{Janssen:2008px}.
The resulting correction to the Friedman equations is suppressed
with respect to the tree level contribution by a factor of
$H^2/m_p^2$ and shows no secular growth.

\hspace{0.3cm}

\vspace{0.3cm}

\begin{flushleft}
PACS numbers: 04.30.-m, 04.62.+v, 98.80.Cq
\end{flushleft}

\vspace{0.1cm}

\begin{flushleft}
$^{*}$ T.M.Janssen@uu.nl, \hspace{1cm} $^{\ddagger}$
T.Prokopec@uu.nl
\end{flushleft}

\end{titlepage}




\section{Introduction}
Since the classic work of 't Hooft and Veltman on quantum gravity
on flat space-time~\cite{'tHooft:1974bx}, many authors have
studied the quantum behavior of gravitons. In particular, because
of the potential relevance for inflationary cosmology, the quantum
behavior of gravitons on a (locally) de Sitter background has been
a widely studied subject over the past
years~\cite{Allen:1986ta,Antoniadis:1986sb,Allen:1986tt,Tsamis:1996qk,Tsamis:1992xa,Tsamis:1996qq,Tsamis:1996qm,Higuchi:2000ge,Higuchi:2000ye,Finelli:2004bm,Christensen:1979iy,Ford:1984hs,Abramo:1996gd,Abramo:1997hu,Abramo:1999wd,Mukhanov:1996ak,Losic:2006ht,Cognola:2005de,Janssen:2007ht,Abramo:1998hj,Abramo:2001dc,Abramo:2001dd,Iliopoulos:1998wq,JanssenProkopecMiao2008,Tsamis:2005je,Parker:1969au}.
Of course de Sitter space is interesting for its relevance for
inflationary cosmology. Therefore there are several works that
study the potential influence of quantum behavior on inflationary
observables~\cite{Janssen:2009pb,Bilandzic:2007nb,Sloth:2006az,Sloth:2006nu}.
Another line of research deals with the back-reaction of
gravitational waves on the background
space-time~\cite{Abramo:1996gd,Abramo:1997hu,Mukhanov:1996ak,Losic:2006ht}.
However of more interest for the present work is the one loop
back-reaction by virtual gravitons on a de Sitter background,
which has been calculated by several authors, using different
techniques~\cite{Finelli:2004bm,Tsamis:2005je,Parker:1969au}.
Since it is not clear whether in these works exactly the same
quantity is calculated and the renormalization schemes differ, the
numerical coefficients differ. However the main result is the
same: one loop graviton contributions to the expectation value of
the energy momentum tensor result in a finite, time independent
shift of the effective cosmological constant. Since the
contribution can always be absorbed in a
counterterm~\cite{Tsamis:2005je}, the exact numerical coefficient
coming from such a calculation has no physical meaning.\\

The goal of this paper is to go beyond the works mentioned above
and calculate the one loop effective action induced by gravitons
in a more general background space-time using dimensional
regularization. The geometry we consider is a
Friedmann-Lema\^{\i}tre-Roberston-Walker (FLRW) geometry with
Hubble parameter $H=\frac{\dot{a}}{a}$ and the additional
constraint that
\begin{equation}\label{eps}
    \epsilon\equiv-\frac{\dot{H}}{H^2}
\end{equation}
is a constant. If the universe contains only one type of matter,
e.g. dust, radiation or a cosmological constant, this constraint
is satisfied. In particular in matter era we have that
$\epsilon=3/2$, in radiation era $\epsilon=2$ and de Sitter space
corresponds to the limit $\epsilon\rightarrow 0$. One immediate
problem with working in such a space-time, instead of de Sitter,
is that for consistency of the Einstein equations the addition of
matter fields is unavoidable. Whereas in de Sitter space, the only
relevant metric fluctuations are the tensor modes (gravitational
waves), in a more general setting one has to take the mixing of
gravitational and matter degrees of freedom into
account~\cite{Brandenberger:2003vk,Mukhanov:1990me,Abramo:1998hj,Abramo:2001dc,Abramo:2001dd,Iliopoulos:1998wq}.\\

Naively one might think that any back-reaction will be
insignificant, since it will in a cosmological setting typically
be suppressed by $H^2/m_p^2$, with $m_p$ the Planck mass. The
reason that this is not necessarily the case has to do with a
second complication. This complication is that gravitons on a
cosmological background space-time behave, apart from the
tensorial structure, somewhat similar to massless scalars, with a
certain amount of coupling to the Ricci scalar. Such a scalar
field however is long known to posses problems in the infrared
~\cite{Ford:1977in,Vilenkin:1982wt}, for a wide class of
cosmological backgrounds. What happens is that due to cosmological
particle production correlations of the Bunch-Davies vacuum grow
too fast, causing the propagator to diverge in the infrared. Such
a divergence should of course be regulated and in a recent work
~\cite{Janssen:2008px} the massless scalar propagator on any
constant $\epsilon$ cosmological space-time was constructed, by
assuming that the universe is described by a spatially compact
manifold. Effectively this simply implies that any momentum
integral has an infrared cut-off at some scale $k_0$. In this
work, we shall use the propagators constructed in
~\cite{Janssen:2008px} to obtain an infrared finite answer.\\
While the answers obtained in this way are infrared finite,
particle production is of course a physical phenomenon. Long range
correlations will therefore still be enhanced as time goes on.
Therefore initially small quantum fluctuations, might -- at least
in principle -- grow significantly in time. The contribution of
these fluctuations to the stress energy tensor might therefore
also grow in time, leading to a potentially significant
back-reaction via the Einstein equations on the background
space-time.

It is precisely this effect, that has led to several papers
concerning quantum contributions to the stress energy tensor in
the case of de Sitter space-time
\cite{Tsamis:1996qk,Tsamis:1996qm,Tsamis:1996qq,Abramo:1998hj,Abramo:2001dc,Abramo:2001dd,Linde:1982uu}.
In these works it is found that at two loop order, the
backreaction due to gravitons might become significantly, while
for scalars this might happen at three loop order.  Such behavior,
if indeed physically viable, has profound implications for the
cosmological constant problem (see {\it e.g.} Ref.~\cite{Nobbenhuis:2004wn}
for a review). In this paper we will not consider these higher
loop effects, but we shall perform a one loop
calculation, but in a more general space-time.\\

In section~\ref{sgeo} we briefly review our background geometry.
In section~\ref{sprop} we briefly summarize the construction of
the propagator as given in~\cite{Janssen:2008px}. In
section~\ref{sprop2} we apply these results to calculate the
relevant kinetic operators and propagators in a model with both
matter and gravitational degrees of freedom. In
section~\ref{soneloop} we calculate the one-loop effective action
contribution to the quantum corrected Friedmann equations and
renormalize the theory. We discuss our results in section
\ref{s_disc} and we conclude in section \ref{s_con}.

\section{Geometry}\label{sgeo}
\begin{figure}
\begin{center}
\includegraphics[width=6in]{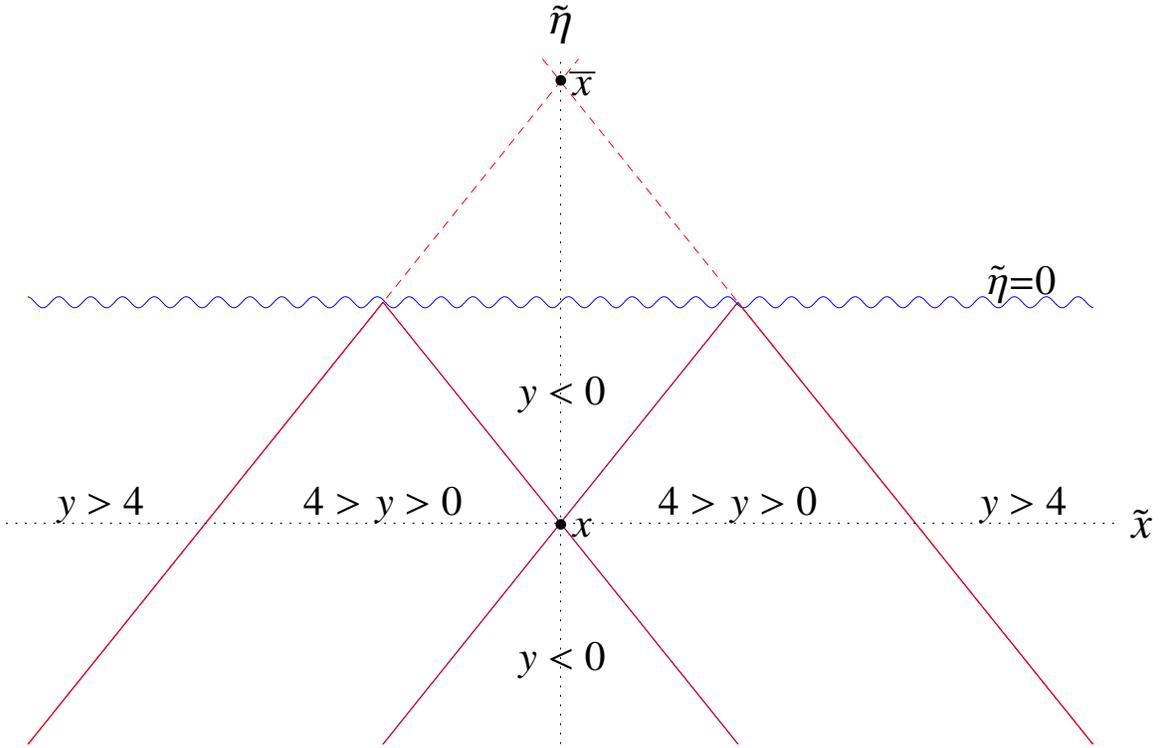}
\caption{The causal structure in the conformal coordinates
(\ref{coord}). The plot assumes $\epsilon<1$, so the coordinates
cover only the region $\eta<0$. The wavy line at $\eta=0$
indicates future infinity. The lightcone of the point $x$ is given
by $y=0$. If $y=4$, the point $\tilde{x}$ lies on the light cone
of an unobservable image charge at the antipodal point
$\bar{x}$.}\label{spacetime}
\end{center}
\end{figure}

The background space-time we consider is is the
Friedmann-Lemaître-Robertson-Walker (FLRW) geometry in conformal
coordinates
\begin{equation}\label{coord}
    g_{\mu\nu}=a^2\eta_{\mu\nu}\qquad;\qquad\eta_{\mu\nu}=\rm{diag}(-1,1,1,1),
\end{equation}
with the additional constraint,
\begin{equation}\label{epsconstr}
    \epsilon\equiv-\frac{\dot{H}}{H^2}=\text{constant}\qquad;\qquad
    H\equiv \frac{\dot{a}}{a}.
\end{equation}
Here a {\it dot} indicates a derivative with respect to cosmological
time $t$, related to the conformal time $\eta$ by $dt=a d\eta$.
The FLRW geometry obeys the Friedmann equations
\begin{equation}\label{fried}
        \frac{3H^2}{\kappa}-\frac{1}{2}\rho_M=0\qquad;\qquad
        -\frac{2\dot{H}}{\kappa}-\frac{1}{2}(\rho_M+p_M)=0
        \,,\qquad
        \kappa=16\pi G_N
        \,,
\end{equation}
with $G_N$ being the Newton constant, $\rho_M$ and $p_M$ are the
energy density and pressure due to matter. If one writes
\begin{equation}
    p_M=w_M\rho_M,
\end{equation}
one immediately finds that (\ref{epsconstr}) implies that $w_M$ is
constant. One can solve $(\ref{fried})$ for $a$ to find
\begin{equation}\label{adingen}
    \begin{split}
        a(\eta)=\Big((\epsilon-1)H_0\eta\Big)^{-1/(1-\epsilon)}\qquad&;\qquad \epsilon=\frac{3}{2}(1+w_M)\\
        H=H_0\Big((\epsilon-1)H_0\eta\Big)^{\epsilon/(1-\epsilon)}=H_0 a^{-\epsilon}\qquad&;\qquad  a\eta=-\frac{1}{1-\epsilon}\frac{1}{H}
    \end{split}.
\end{equation}
Notice that if $\epsilon<1$, $\eta$ is negative and the expansion
of the universe is accelerating. If $\epsilon>1$, $\eta$ is
positive and the expansion is decelerating. $H_0$ is chosen such
that the $\epsilon\rightarrow 0$ limit of $H$ corresponds to the
one given in~\cite{Janssen:2007ht}. Given two points $x$ and
$\tilde{x}$, the following distance function will prove to be
useful
\begin{equation}\label{y}
    y(x;\tilde{x})=\frac{\Delta x^2(x;\tilde{x})}{\eta\tilde{\eta}}
    =\frac{1}{\eta\tilde{\eta}}(-(|\eta-\tilde{\eta}|
    -\imath\varepsilon)^2+||\vec{x}-\vec{\tilde{x}}||^2)
\,.
\end{equation}
Here the infinitesimal $\varepsilon>0$ refers to the Feynman
(time-ordered) pole prescription. In the following we shall omit
the explicit argument of $y(x;\tilde{x})$, since this will be
clear from the context. In de Sitter space $y$ is related to the
geodesic distance $l$ as $y=4\sin^2(\frac{1}{2}H l)$. If $y<0$,
points $\tilde{x}$ are timelike related to $x$, and if $y>0$, they
are spacelike related. We define the antipodal point $\bar{x}$ of
$x$ by the map $\eta\rightarrow -\eta$~\cite{Allen:1985ux}. Notice
that, since as long as $\epsilon$ is constant, $\eta$ is either
positive or negative, this point is not covered by our coordinate
patch. If $y=4$, $\tilde{x}$ lies on the lightcone of an
(unobservable) image charge at the antipodal point $\bar{x}$, see
figure \ref{spacetime}.

The Levi-Civit\`a connection is,
\begin{equation}
 \Gamma^\alpha_{\mu\nu} = \frac{a'}{a}
        \left[
             \delta^\alpha_{\;\mu}\delta_{\nu}^{\;0}
            + \delta^\alpha_{\;\nu}\delta_{\mu}^{\;0}
            + \delta^\alpha_{\;0}\eta_{\mu\nu}
        \right]
\,,
\label{LCconnection}
\end{equation}
while the curvature tensors are given by
\begin{equation}
    \begin{split}
        R^\alpha{}_{\mu\beta\nu}=&\Big(\frac{a''}{a}-2\Big(\frac{a'}{a}\Big)^2\Big)
        \Big(\delta^\alpha_\nu\delta^0_\mu\delta^0_\beta-\delta^\alpha_0\delta^0_\nu\eta_{\mu\beta}
        -\delta^\alpha_\beta\delta^0_\nu\delta^0_\mu+\delta^0_\beta\delta^\alpha_0\eta_{\mu\nu}\Big)
        \\
        &-\Big(\frac{a'}{a}\Big)^2
        \Big(\delta^\alpha_\nu\eta_{\mu\beta}-\delta^\alpha_\beta\eta_{\mu\nu}\Big)\\
        R_{\mu\nu}=&\Big(\frac{a''}{a}-2\Big(\frac{a'}{a}\Big)^2\Big)\Big(\eta_{\mu\nu}
        -(D-2)\delta_\mu^0\delta_\nu^0\Big)+\Big(\frac{a'}{a}\Big)^2(D-1)\eta_{\mu\nu}\\
        R=&\Big(\frac{a''}{a^3}-2\Big(\frac{a'}{a^2}\Big)^2\Big)2(D-1)
        +\Big(\frac{a'}{a^2}\Big)^2D(D-1)
\,,
    \end{split}
    \label{curvature invariants}
\end{equation}
where $a' = da/d\eta$ and $D$ denotes the number of space-time dimensions.

\section{Scalar Propagator}\label{sprop}
Since it will turn out that the graviton propagator can be
expressed in terms of massless scalar propagators with different
amount of conformal coupling, we need to find a solution for the
following scalar Klein-Gordon equation
\begin{equation}\label{Scal_KG}
    \sqrt{-g}\left(\Box-\xi
    R\right)\imath\Delta(x;\tilde{x})=\imath\delta^D(x-\tilde{x})\,,
\end{equation}
where $\xi$ is a dimensionless scalar field coupling to the Ricci
curvature scalar, and $\Box$ denotes the d'Alembertian, which
reads when acting on a scalar
\begin{equation}
    \sqrt{-g}\Box\phi =
    \partial_\mu\sqrt{-g}g^{\mu\nu}\partial_\nu\phi.
\end{equation}
If $\epsilon$ is constant we can rewrite this equation as follows
\begin{equation}
   \Big[\eta^{\mu\nu}\partial_\mu\partial_\nu+\f{1}{\eta^2}\Big(\nu^2-\f{1}{4}\Big)\Big]\Big[(a\tilde{a})^{\f{D}{2}-1}\imath\Delta(x;\tilde{x})\Big]
  =\imath\delta^D(x-\tilde{x}),
\end{equation}
where $\tilde{a}=a(\tilde{x})$ and

\begin{equation}\label{scalnuorig}
    \nu^2=\Big(\frac{D-1}{2}\Big)^2-(1-\epsilon)^{-2}\Big((D-1)(D-2\epsilon)\xi
                     -\frac{1}{2}(D-1)(D-2)\epsilon+\frac{D}{4}(D-2)\epsilon^2\Big).
\end{equation}
The properties of such a scalar field propagator on a de Sitter
background ($\epsilon=0$ in our notation) have been studied in the
past \cite{Allen:1985ux,Mottola:1984ar,Chernikov:1968zm}.

This propagator in a general constant $\epsilon$ space is given in
\cite{Janssen:2008px}, by integrating over the mode functions and
we simply quote the result here. The massless scalar propagator is
infrared divergent if the index $\nu$ (which is defined in
(\ref{scalnuorig}) is less then $(D-1)/2$. Therefore we work on a
compact spatial manifold, with comoving radius $k_0^{-1}$. In this
case, the integral over the mode functions becomes a discrete sum.
However, if $k_0$ is small enough, it is generally valid to
approximate this sum by an integral, with an infrared cutoff. This
approach has passed sever consistency checks on de Sitter
space\cite{Tsamis:1996qk,MW,Kahya:2007bc}. In the final propagator
we can then recognize two different types of contributions: one
coming from the 'infinite volume' integral (thus the integral,
without an infrared cut-off), and one from the subtraction of the
lower range of the integral.

\begin{equation}\label{final_cutoff_prop}
    i \Delta(x;\tilde{x})=i \Delta_\infty(x;\tilde{x})+\sum_{N=0}^\infty
     \delta i\Delta_N(x;\tilde{x})+\sum_{N=0}^\infty
     \delta i\Delta^N(x;\tilde{x})
\, .
\end{equation}
Here the infinite volume propagator is
\begin{eqnarray}
i\Delta_{\infty}(x;\tilde{x}) = \frac{\Bigl[ (1\!-\!\epsilon)^2 H
\tilde{H} \Bigr]^{\frac{D}2 -1} }{(4 \pi)^{\frac{D}2}}
\frac{\Gamma(\frac{D-1}2 \!+\!\nu) \Gamma(\frac{D-1}2 \!-\!
\nu)}{\Gamma(\frac{D}2)}
\; \mbox{}_2 F_1\biggl(\frac{D\!-\!1}2 \!+\! \nu, \frac{D\!-\!1}2
\!-\! \nu ; \frac{D}2 ; 1 \!-\! \frac{y}4\biggr)
\nonumber\\
 \label{Tomislav}
\end{eqnarray}
and the correction terms due to the infrared divergence are
\begin{equation}\label{corrections2}
\begin{split}
\delta i \Delta_N(x;\tilde{x})&=
-\f{\Big(H\tilde{H}(1-\epsilon)^2\Big)^{\f{D}{2}-1}}{(4\pi)^{\f{D}{2}}}\f{2
z_0^{2N+(D-1)-2\nu}}{2N+(D-1)-2\nu}\\
&\qquad\qquad\times\f{\Gamma(2\nu)\Gamma(\nu)}{\Gamma(\f{1}{2}+\nu)\Gamma(\f{D-1}{2})}
\sum_{k=0}^{N} \sum_{\ell = 0}^{N-k} a_{k\ell}
\Big(\f{r}{\tilde{\eta}}\Big)^{2k}
\Big(\f{\eta}{\tilde{\eta}}\Big)^{2\ell-N}\\
\delta i \Delta^N(x;\tilde{x})&
 =\delta i \Delta_N(x;\tilde{x})|_{\nu\leftrightarrow -\nu}
\,,
\end{split}
\end{equation}
where $z_0=k_0|\eta|$ and
\begin{equation}
a_{k\ell} = \Bigl(-\frac1{4}\Bigr)^N  \frac1{k! \, \ell! \, (N
\!-\! k \!-\! \ell)!} \, \frac{\Gamma(\frac{D-1}2) \, \Gamma^2(1
\!-\! \nu)}{ \Gamma(k \!+\! \frac{D-1}2) \Gamma(\ell \!+\! 1 \!-\!
\nu) \Gamma(N \!-\! k \!-\! \ell \!+\! 1 \!-\! \nu)} \; .
\label{akl}
\end{equation}

Notice that in an accelerating space-time ($\epsilon<1$) $\eta$
approaches  zero at late times, while in a decelerating space-time
($\epsilon>1$), $\eta$ approaches infinity. This is an interesting
observation, since it implies that in an accelerating universe,
$i\Delta_N$ contains terms that grow in time if $\nu<\f{D-1}{2}$.
In a decelerating universe both $i\Delta_N$ and $i\Delta^N$ can
contain growing terms. It is precisely these growing terms that
can lead to a secular growth of quantum effects, described in the
introduction. There are also other regularizations possible for
the infrared. For example in~\cite{JanssenProk} the infrared is
regulated by matching an infrared finite space-time to the
space-time under consideration. This ensures that no infrared
divergences can occur. Results calculated using this
regularization are similar in accelerating space-times, but show
differences in a decelerating space-time.

\section{The gravitational and matter propagators}\label{sprop2}
For our model we shall consider the action of gravity plus a
scalar field $\hat{\phi}=\hat{\phi}(x)$ with an arbitrary
potential $V(\hat{\phi})$
\begin{equation}
    S=\int\sqrt{-\hat{g}}\Big(\f{\hat{R}-(D-2)\Lambda}{\kappa}
            -\f{1}{2}(\partial\hat{\phi})^2-V(\hat{\phi})\Big),
\end{equation}
where $\kappa=16\pi G_N = 16\pi/m_{\rm P}^2$ denotes the
(rescaled) Newton constant, $m_{\rm P} \simeq 1.2\times
10^{19}~{\rm GeV}$ is the Planck mass, $\Lambda$ denotes the
cosmological constant, $D$ is the number of space-time
dimensions and $(\partial\hat{\phi})^2
 = \hat g^{\mu\nu}(\partial_\mu\hat{\phi})(\partial_\nu\hat{\phi})$.
By an appropriate choice of the potential $V$, such a model can
mimic any mixture of fluids which are relevant for the evolution
of the univese. We consider a conformally flat FLRW background and
split the fields in a background contribution and a quantum
contribution~\cite{Tsamis:1992xa}\cite{Iliopoulos:1998wq}
\begin{equation}
\begin{split}
    \hat{g}_{\mu\nu}&=g_{\mu\nu}(\eta) + \delta g_{\mu\nu}
                     = a^2(\eta_{\mu\nu}+\sqrt{\kappa}\psi_{\mu\nu})\\
    \hat{g}^{\mu\nu}&= g^{\mu\nu}(\eta) + \delta g^{\mu\nu}
                     =  a^{-2}(\eta^{\mu\nu}
    -\sqrt{\kappa}a^4\psi^{\mu\nu})+\mathcal{O}(\psi^2)\\
    \hat{\phi}&=\Phi(\eta)+\phi\,,
\end{split}
\end{equation}
where $\delta g_{\mu\nu} \equiv h_{\mu\nu}$ denotes the graviton field,
$\psi_{\mu\nu}$ is the pseudo-graviton field and
$\delta g^{\mu\nu} = - h^{\mu\nu} + h^\mu_{\;\alpha}h^{\alpha\nu}
+ {\cal O}(h^3)$.
Notice that indices on the pseudo-graviton field $\psi_{\mu\nu}$
are raised and lowered with the full background metric
$g_{\mu\nu}(\eta)= a^2\eta_{\mu\nu}$. The
background scalar field $\Phi$ is homogeneous and thus only depends on
(conformal) time. The background fields obey the tree level
Friedmann equations and the scalar field equation of motion:
\begin{equation}\label{backgrond1}
    \begin{split}
    H^2-\f{1}{D-1}\Lambda-\f{\kappa}{(D-1)(D-2)}\Big(\f{1}{2a^2}\Phi'^2+V(\Phi)\Big)&=0\\
    a^{-1}H'+\f{D-1}{2}H^2-\f{1}{2}\Lambda+\f{\kappa}{2(D-2)}\Big(\f{1}{2a^2}\Phi'^2-V(\Phi)\Big)&=0\\
    \Phi''+(D-2)aH\Phi'+a^2\f{\partial V}{\partial
    \phi}(\Phi)&=0
\,,
    \end{split}
\end{equation}
from which one can derive the following identities
\begin{equation}\label{backgrond2}
    \begin{split}
        \sqrt{\kappa}\Phi'=\sqrt{2(D-2)\epsilon}aH\;;&
       \qquad \sqrt{\kappa}\Phi''
             =\sqrt{2(D-2)\epsilon}(1-\epsilon)a^2  H^2 +\mathcal{O}(\epsilon')\\
    \sqrt{\kappa}\f{\partial V}{\partial
    \phi}(\Phi)=-\sqrt{2(D-2)\epsilon}(D-1-\epsilon)H^2\;;
            &\qquad  \f{\partial^2 V}{\partial
    \phi^2}(\Phi)=2(D-1-\epsilon)\epsilon H^2
    +\mathcal{O}(\epsilon').
    \end{split}
\end{equation}
Notice that in slow roll inflation $\epsilon'$ is non zero. In
order to facilitate a comparison, we give here the slow roll
expressions for the second and fourth relation (the first and
third do not change)
\begin{equation}\label{backgrond6}
    \begin{split}
\sqrt{\kappa}\Phi''
             &=\sqrt{\f{2\epsilon_V}{D-2}}a^2  H^2\Big(D+(D-4)\epsilon_V-\f{(D-1)(D-2)}{2}\eta_V\Big)\\
\f{\partial^2 V}{\partial
    \phi^2}(\Phi)&=\f{(D-1)(D-2)}{2}\eta_V H^2,
    \end{split}
\end{equation}
in terms of the slow roll parameters
\begin{equation}
    \epsilon_V\equiv \f{1}{\kappa}\Big(\f{\partial V}{V\partial
    \phi}\Big)^2\qquad;\qquad\eta_V\equiv\f{2}{\kappa}\f{\partial^2 V}{V\partial
    \phi^2}.
\end{equation}
Notice that in the slow roll approximation $\epsilon$ and
$\epsilon_V$ are identical.

For the purpose of this paper we are only interested in the
quadratic perturbations in the fields. After many partial
integrations we find~\footnote{An analogous result can be found in
                               Ref.~\cite{Iliopoulos:1998wq}. The main difference is
             that our result~(\ref{L2}) includes also terms that vanish on-shell.}
\begin{equation}\label{L2}
    \begin{split}
    \mathcal{L}^{(2)}=&a^{D+4}\psi^{\mu\nu}\Bigg(\Big(\Box_s-\mathcal{W}\Big) \Big(\f{1}{4}\delta^\rho_\mu\delta^\sigma_\nu-\f{1}{8}\eta_{\mu\nu}\eta^{\rho\sigma}\Big)+\mathcal{X}\delta_\mu^0\delta^\rho_\nu\delta^\sigma_0-\mathcal{Y}\eta_{\mu
                0}\delta_\nu^0\eta^{\rho\sigma}\Bigg)\psi_{\rho\sigma}\\
                &-a^{D-2}\psi_{00}\Big(\sqrt{\kappa}\Phi''\Big)\phi+a^{D-2}\sqrt{\kappa}\eta^{\mu\nu}\psi_{\mu\nu}\mathcal{Z}\phi\\
                &+\f{1}{2}a^D\phi\Big(\Box_s-\f{\partial^2
    V}{\partial\phi^2}(\Phi)+a^{-2}\Phi'^2\kappa\Big)\phi
       +\f{1}{2}\sqrt{-g}g^{\alpha\beta}F_\alpha F_\beta
\,,
    \end{split}
\end{equation}
where we defined
\begin{equation}\label{FWXYZ}
    \begin{split}
        F_\alpha&=a^{2}\nabla_\mu\Big(\psi^\mu_\alpha
                -\f{1}{2}\delta^\mu_\alpha g^{\rho\sigma}\psi_{\rho\sigma}\Big)
                 -\phi\Phi'\sqrt{\kappa}\delta_\alpha^0\\
        \mathcal{W}&=2(D-2)
                     \bigg[a^{-1}H'+\f{D-1}{2}H^2-\f{1}{2}\Lambda
            +\f{\kappa}{2(D-2)}\Big(\f{1}{2a^2}\Phi'^2-V(\Phi)\Big)\bigg]\\
        \mathcal{X}&=\Big(\f{1}{2}\kappa  a^{-2}\Phi'^2-\f{D-2}{2}(H^2-a^{-1}H')\Big)\\
        \mathcal{Y}&=\Big(\f{1}{4}\kappa a^{-2}\Phi'^2+\f{D-2}{2}a^{-1}H'\Big)\\
        \mathcal{Z}&=-\f{1}{2}\bigg(\Phi''+(D-2)aH\Phi'+\f{\partial
                V}{\partial\phi}(\Phi)a^2\bigg)
    \end{split}
\end{equation}
and
\begin{equation}
\Box_s=
\f{1}{\sqrt{-g}}\partial_\mu\sqrt{-g}g^{\mu\nu}\partial_\nu
\end{equation}
is the d'Alembertian as it acts on a scalar field. We add a gauge
fixing term
\begin{equation}\label{gauge}
    \mathcal{L}_{GF}=-\f{1}{2}\sqrt{-g}g^{\alpha\beta}F_\alpha
    F_\beta
\end{equation}
and therefore we also need to add a ghost lagrangian
\begin{equation}
    \mathcal{L}_{\rm ghost}=-\sqrt{-g}a^{-2}\bar{V}^\mu\delta F_\mu
\,,
\end{equation}
where we consider the change of $F_\mu$ under infinitesimal
coordinate transformations
$x'^\mu = x^\mu+\sqrt{\kappa}V^\mu$.
From $\phi(x^\mu) =\phi({x'}^\mu) + \delta \phi$
and $\psi_{\mu\nu}(x^\mu) =\psi_{\mu\nu}({x'}^\mu)
     + \delta \psi_{\mu\nu}$, we find up to first order in $V^\mu$:
\begin{equation}
    \begin{split}
        \delta\phi&=-\sqrt{\kappa}V^0\Phi'\\
        \delta \psi_{\mu\nu}&=-a^{-2}\Big(g_{\alpha\nu}\partial_\mu
V^\alpha+g_{\alpha\mu}\partial_\nu
V^\alpha+2\Big(\frac{a'}{a}\Big)g_{\mu\nu} V^0\Big)
    \end{split}
\end{equation}
and thus
\begin{equation}\label{ghost_lag}
   \mathcal{L}_{\rm ghost}
           =a^D\eta_{\alpha\beta}\bar{V}^\alpha
            \Bigg(\delta^\beta_\mu\Box_s
               -(D-2)(H^2-a^{-1}H')\delta^\beta_0\delta_\mu^0
               +a^{-2}\kappa\Phi'^2\delta^\beta_0\delta^0_\mu
            \Bigg)
        V^\mu
\,,
\end{equation}
where $V$ and $\bar{V}$ are the ghost and anti-ghost fields.

The quadratic lagrangian~(\ref{L2}--\ref{FWXYZ}) contains mixing
between the different components. The following field redefinition
removes the mixing between $\psi_{ij}$ and $\psi_{00}$ and $\phi$
\emph{on-shell} \footnote{In the special case when $D=4$, this and
Eq.~(\ref{Rotation}) below agree with
Ref.~\cite{Iliopoulos:1998wq}.}.
\begin{equation}\label{fieldredef}
    \begin{split}
        \psi_{ij}&=z_{ij}+\f{\delta_{ij}}{D-3}z_{00}\\
        \psi_{00}&=z_{00}\\
        \psi_{0i}&=z_{0i}
\,.
    \end{split}
\end{equation}
The resulting quadratic lagrangian (which is still valid off
shell) can be written as
\begin{equation}\label{lagofshell}
\mathcal{L}^{(2)}+  \mathcal{L}_{GF} + \mathcal{L}_{\rm ghost}
     =\f{1}{2}X^T_{ij}G^{ijkl}X_{kl}+\f{1}{2}z_{0i}\mathcal{D}^{ij}_{\rm vector}z_{oj}
      +\bar{V}^\alpha\mathcal{D}^{\rm ghost}_{\alpha\beta}V^\beta,
\end{equation}
where
\begin{equation}\label{kinetic}
    \begin{split}
        X_{ij}&=\left(\begin{array}{c}
                        z_{ij}\\
                        z_{00}\\
                        \phi\\
                      \end{array}\right)\\
        G^{ijkl}&=\left(\begin{array}{ccc}
                    \mathcal{D}_{\rm tensor}^{ijkl} &a^{D}\mathcal{Y}\delta^{ij}&a^{D-2}\sqrt{\kappa}\mathcal{Z}\delta^{ij}\\
                    a^{D}\mathcal{Y}\delta^{kl}
                  &\mathcal{D}_{\rm scalar}&a^{D-2}\sqrt{\kappa}(\f{2}{D-3}\mathcal{Z}-\Phi'')\\
                    a^{D-2}\sqrt{\kappa}\mathcal{Z}\delta^{kl}
         &a^{D-2}\sqrt{\kappa}(\f{2}{D-3}\mathcal{Z}-\Phi'')&\mathcal{D}_\phi\\
                    \end{array}\right)\\
        \mathcal{D}_{\rm tensor}^{ijkl}&=a^D\Big(\Box_s-\mathcal{W}\Big)\Big(\f{1}{2}\delta^{ik}\delta^{jl}-\f{1}{4}\delta^{ij}\delta^{kl}\Big)\\
        \mathcal{D}_{\rm vector}^{ij}&=-a^D\Bigg(\Big(\Box_s-\mathcal{W}\Big)+2\mathcal{X}\Bigg)\eta^{ij}\\
        \mathcal{D}_{\rm scalar}&=a^D\Bigg(\f{D-2}{2(D-3)}\Big(\Box_s-\mathcal{W}\Big)+2\mathcal{X}+\frac{4}{D-3}\mathcal{Y}\Bigg)\\
        \mathcal{D}_\phi&=a^D\Big(\Box_s-\f{\partial^2
        V}{\partial\phi^2}(\Phi)+a^{-2}\Phi'^2\kappa\Big)\\
        \mathcal{D}^{\rm ghost}_{\alpha\beta}&=a^D\Big(\eta_{\alpha\beta}\Box_s+2\mathcal{X}\delta_\alpha^0\eta_{\beta
        0}\Big).
    \end{split}
\end{equation}
Note that $G^{ijkl}$ contains the tensor as well as
the two scalar (gravitational and matter) kinetic operators.

\subsection{The propagators}
We shall now construct the propagators, associated to the various
modes in (\ref{lagofshell}). In general this will not be possible,
due to the nontrivial dependence of the propagators on the
background fields and the mixing between the different modes.
Therefore we shall restrict ourselves in calculating the
\emph{on-shell} propagators. As will be clear from the discussion
in section \ref{soneloop} this will be sufficient to calculate the
one loop effective action. It will turn out that kinetic operators
can be written in terms of
\begin{equation}\label{general_kinetic}
        \mathcal{D}_n\equiv\sqrt{-g}
                \Big[\Box_s-n\Big(D-n-1+\f{n(n-1)}{2}\epsilon\Big)(1-\epsilon)H^2\Big]
\qquad (n=0,1,2)\,,
\end{equation}
with an associated propagator
\begin{equation}
        \mathcal{D}_ni\Delta_n(x;\tilde{x})=i\delta^D(x-\tilde{x})
\qquad (n=0,1,2)\,.
\end{equation}
The operator (\ref{general_kinetic}) is however nothing but the
kinetic operator for the massless scalar field, with conformal
coupling
\begin{equation}
    \xi =
    \f{n(D-n-1+\f{n(n-1)}{2}\epsilon)}{(D-1)(D-2\epsilon)}(1-\epsilon)
\end{equation}
such that we can use the propagators calculated in the previous
sections with the parameter $\nu$, given in (\ref{scalnuorig})
replaced by
\begin{equation}\label{scalnun}
    \nu_{n}^2=\Big(\frac{D-1-\epsilon}{2(1-\epsilon)}\Big)^2-
    \frac{n\Big(D-n-1+\f{n(n-1)}{2}\epsilon\Big)
}{1-\epsilon} \,.
\end{equation}
We find for the kinetic operators for the vector and the ghost
\emph{on-shell}
\begin{equation}\label{props}
    \begin{split}
        \mathcal{D}_{\rm vector}^{ij}\left|_{\rm on\; shell}\right.&=-\mathcal{D}_1\delta^{ij}\\
        \mathcal{D}^{\rm ghost}_{\mu\nu}\left|_{\rm on\; shell}\right.&=\Big(\bar{\eta}_{\mu\nu}\mathcal{D}_0+\delta^0_\mu\eta_{\nu_0}\mathcal{D}_1\Big)
    \end{split}
\end{equation}
 and their associated propagators:
\begin{equation}\label{props1}
    \begin{split}
        i{}_j\Delta_{\rm vector}^k&=-\delta_j^ki\Delta_1\\
        i{}_\alpha \Delta_{\rm ghost}^\rho
   &=\Big(\bar{\delta}^\rho_\alpha i\Delta_0
             +\delta_\alpha^0\delta^\rho_0i\Delta_1\Big).
    \end{split}
\end{equation}
There is still  mixing between $z_{00}$ and $\phi$. The
\emph{on-shell} part of the mixing can be removed by the following
rotation
\begin{equation}\label{Rotation}
    \begin{split}
        X&=R Y\\
        Y_{ij}&=\left(\begin{array}{c}
            z_{ij}\\
            \chi\\
            \nu\\
                \end{array}\right)\\
        R&=\left(\begin{array}{ccc}
        1&0&0\\
        0&\sqrt{\lambda}\cos(\theta)&-\sqrt{\lambda}\sin(\theta)\\
        0&\f{1}{\sqrt{\lambda}}\sin(\theta)&\f{1}{\sqrt{\lambda}}\cos(\theta)\\
         \end{array}\right)\\
         \lambda&=\sqrt{\f{2(D-3)}{D-2}}\\
         \tan(2\theta)&= \frac{2\sqrt{(D-3)\epsilon}}{D-3-\epsilon}
 \,,\qquad \theta=\arccos\Big(-\sqrt{\f{\epsilon}{D-3+\epsilon}}\Big).
    \end{split}
\end{equation}
This rotation reduces the \emph{on-shell} part of the term
$\frac{1}{2}X^T G X$ to
\begin{equation}
    \frac{1}{2}Y_{ij}^T G^{ijkl}_{diag}Y_{kl}\left|_{\rm on\; shell}\right.=
    \frac{1}{2}Y_{ij}^T\left(\begin{array}{ccc}
                    \Big(\frac{1}{2}\delta^{ki}\delta^{lj}-\frac{1}{4}\delta^{ij}\delta^{kl}\Big)\mathcal{D}_0
                    &0&0\\
                    0&\f{1}{\lambda}\mathcal{D}_0&0\\
                    0&0&\f{1}{\lambda}\mathcal{D}_2\\
                    \end{array}\right) Y_{kl}\,.
\end{equation}
Thus the associated propagator matrix $\mathcal{M}_{diag}$,
defined by
\begin{equation}
        G_{diag}\mathcal{M}_{diag}(x;\tilde{x})\left|_{\rm on\; shell}\right.
               =\mathbf{1}\delta^D(x-\tilde{x})
\end{equation}
is given by
\begin{equation}
    \begin{split}
\mathcal{M}_{diag}\left|_{\rm on\; shell}\right.
           &=\left(\begin{array}{ccc}
                    {}_{rs}\Delta_{kl}&0&0\\
                    0&\lambda\Delta_0&0\\
                    0&0&\lambda\Delta_2\\
                    \end{array}\right)\\
        i{}_{rs}\Delta_{kl}&=\Big(2\delta_{r(k}\delta_{l)s}
           -\frac{2}{D-3}\delta_{rs}\delta_{kl}\Big)i\Delta_0
\,.
        \end{split}
\end{equation}
To avoid confusion with the subscript $diag$, we have omitted the
explicit Lorentz indices in $\mathcal{M}_{diag}$ and $G_{diag}$.
It now follows that the non-diagonal propagator matrix
${}_{rs}\mathcal{M}_{kl}$ that inverts $G^{ijkl}$ \emph{on-shell}
is
\begin{equation}\label{props2}
    \begin{split}
    {}_{rs}\mathcal{M}_{kl}\left|_{\rm on\; shell}\right.
    &=R\mathcal{M}_{diag}R^T\left|_{\rm on\; shell}\right.
\\
    &=\left(\begin{array}{ccc}
            {}_{rs}\Delta_{kl}&0&0\\
            0&\lambda^2(\cos^2(\theta)\Delta_0+\sin^2(\theta)\Delta_2)
            &\lambda\cos(\theta)\sin(\theta)(\Delta_0-\Delta_2)\\
            0&\lambda\cos(\theta)\sin(\theta)(\Delta_0-\Delta_2)&
            (\sin^2(\theta)\Delta_0+\cos^2(\theta)\Delta_2)\\
            \end{array}\right)\,.
            \end{split}
\end{equation}
This finishes the construction of all the propagators and we
indeed find that all modes can be described in terms of
$\Delta_n$.

 \subsection{One-loop effective action}\label{soneloop}

 In this section we shall first sketch using a simple example how
to calculate the correction to the Friedmann equation due to the
one-loop effective action. Afterwards we shall apply it to the
case at hand. We consider as an example a model with an action
\begin{equation}
    S = S_0 + S_\chi,
\end{equation}
where $\chi$ is a quantum scalar field with an action
$S_\chi=S_\chi[\chi]$ and $S_0$ is the classical action of any
background fields (including for example the Einstein-Hilbert
action). For any action $S_\chi$ that is quadratic in $\chi$,
\begin{equation}
    S_\chi=\int d^Dx \f{1}{2}\chi \mathcal{D}_\chi \chi,
\end{equation}
one gets the following effective action
\begin{equation}
    \Gamma=S_0-i\ln\Big(\f{1}{\sqrt{\rm{Det}(\mathcal{D}_\chi)}}\Big)=S_0+\f{i}{2}{\rm{Tr}}\ln\big(\mathcal{D}_\chi\big).
\end{equation}
Here the trace involves tracing over the Lorentz indices and
space-time integration of the operator at coincidence
~\cite{Birrell:1982ix} and $\mathcal{D}_\chi$ is the kinetic
operator of the field.

 While in principle one could -- at least formally -- evaluate the effective
action, the object one is eventually interested in is the
effective Friedmann equation, {\it i.e.} the equations of motion
associated with the background metric. Moreover in the present
case we need to work under the constraint that $\epsilon$ is
constant. As long as $\dot \epsilon$ remains small, there is no
problem with imposing such a constraint on the equations of
motion. On the other hand, imposing this constraint on the level
of the action typically changes the dynamics substantially.
Therefore we shall not attempt to explicitly construct the
effective action, but instead we shall directly calculate the
effective Friedmann equation. By taking the functional derivative
of the action with respect to the scale factor $a=a(\eta)$, we
obtain the Einstein trace equation, that is the $-(00)+(D-1)(ii)$
component of the Einstein equation. The second Friedmann equation
can then always be obtained by imposing the Bianchi identity. Thus
we are interested in calculating
\begin{equation}\label{effectex}
    \begin{split}
        \frac{\delta \Gamma}{\delta a}&=\frac{\delta S_0}{\delta a}+\f{i}{2}\frac{\delta }{\delta a}{\rm{Tr}}\ln\big(\mathcal{D}_\chi\big)\\
        &=Va^{D\!-\!1}\bigg[\frac{D(D\!-\!1)(D\!-\!2)}{\kappa}
          \Big(H^2-\frac{1}{D\!-\!1}\Lambda+\frac2{D} a^{-1}H'\Big)
         + (D\!-\!1)p_M-\rho_M \bigg]\\
        &\hskip 1cm
        +\f{i}{2}\frac{\delta }{\delta a}{\rm{Tr}}\ln\big(\mathcal{D}_\chi\big)
        \,,
    \end{split}
\end{equation}
where $V=\int d^{D-1}x$ denotes the volume of space and we assumed
that $S_0$ contains the Einstein-Hilbert action, and matter fields
with an associated total pressure and energy $p_M$ and $\rho_M$.
Notice that the quantum contribution $\f{i}{2}\frac{\delta
}{\delta a}{\rm{Tr}}\ln\big(\mathcal{D}_\chi\big)$ is by
definition nothing but $V a^{D-1} g^{\mu\nu}\langle
T_{\mu\nu}\rangle$, where $\langle T_{\mu\nu}\rangle$ is the one
loop expectation value of the stress-energy
tensor~\cite{Birrell:1982ix}. We now consider the calculation of
this contribution. To be explicit, we shall assume that $\chi$ is
a massless minimally coupled scalar and therefore
\begin{equation}
    \mathcal{D}_\chi=\sqrt{-g}\Box
\end{equation}
with an associated propagator $i\Delta(x;\tilde{x})$ which obeys
\begin{equation}\label{chi:KGeq}
    \mathcal{D}_\chi i\Delta(x;\tilde{x})=i\delta^D(x-\tilde{x})
\,.
\end{equation}
Instead of considering the $\chi$ field, it is convenient to use a
rescaled field
\begin{equation}\label{chi:rescale}
    \hat{\chi} = a^{\f{D}{2}-1}\chi
\end{equation}
with an associated kinetic operator and propagator
\begin{equation}\label{chi:prop}
    \begin{split}
        \hat{\mathcal{D}}_\chi&=\partial^2+\f{(D-2)(D-4)}{4}\f{a'^2}{a^2}+\f{D-2}{2}\f{a''}{a}\\
        i\hat{\Delta}(x;\tilde{x})&=a^{\f{D}{2}-1}\hat{a}^{\f{D}{2}-1}i\Delta(x;\tilde{x}).
        \end{split}
\end{equation}
We shall see explicitly later that -- up to a $D$ dimensional
divergent delta function that does not contribute in dimensional
regularization -- we have that
\begin{equation}
   \frac{\delta}{\delta a} {\rm{Tr}}\ln\big(\mathcal{D}_\chi\big)
  =  \frac{\delta}{\delta a} {\rm{Tr}}\ln\big(\mathcal{\hat{D}}_\chi\big)
\,.
\end{equation}
We shall now show how to calculate the trace logarithm
contribution to (\ref{effectex}) due to the field $\hat{\chi}$. To
be precise we give the exact coordinate dependence of each term
indicated by  $x^\mu$, $y^\mu$, and $z^\mu$
\begin{equation}\label{example1}
    \begin{split}
        \f{i}{2}\frac{\delta }{\delta a(z^0)}{\rm{Tr}}\ln\big(\hat{\mathcal{D}}_\chi(x)\big)&=\f{i}{2}{\rm{Tr}}\Big(\hat{\Delta}(x;y)\Big(\frac{\delta }{\delta
        a(z^0)}\hat{\mathcal{D}}_\chi(x)\Big)\Big)\\
        &=\f{i}{2}\int d^D x\,\int d^Dy
        \hat{\Delta}(x;y)\delta^D(x-y)\\
        &\qquad\Bigg[\Big(-\f{(D-2)(D-4)}{2}\f{a'(x^0)^2}{a(x^0)^3}-\f{D-2}{2}\f{a''(x^0)}{a(x^0)^2}\Big)\delta(x^0-z^0)\\
        &\qquad+\f{(D-2)(D-4)}{2}\f{a'(x^0)}{a(x^0)^2}\partial_{x^0}\delta(x^0-z^0)+\f{D-2}{2a(x^0)}\partial_{x^0}^2\delta(x^0-z^0)\Bigg]\\
        &=\f{1}{2}V \Bigg[
        \Big(-\f{(D\!-\!2)(D\!-\!4)}{2}\f{a'(z^0)^2}{a(z^0)^3}-\f{D\!-\!2}{2}\f{a''(z^0)}{a(z^0)^2}\Big)i\hat{\Delta}(z^0;z^0)\\
        &-\f{(D\!-\!2)(D\!-\!4)}{2}\partial_{z^0}
        \Big(\f{a'(z^0)}{a(z^0)^2}i\hat{\Delta}(z^0;z^0)\Big)+\f{D\!-\!2}{2}\partial_{z^0}^2\Big(\f{1}{a(z^0)}i\hat{\Delta}(z^0;z^0)\Big)\Bigg]\\
        &=-\f{D-2}{4}V a^{D-1}\Box_z i\Delta(z;z)
\,,
    \end{split}
\end{equation}
where in going from the second to the third step we used the delta
functions to change  $\partial_{x^0}$ to $-\partial_{z^0}$ and we
used that the propagators at coincidence are a function of time
only. In the last step we used (\ref{chi:prop}) to rewrite
$i\hat{\Delta}$ in terms of $i\Delta$.

Since the trace of the one loop expectation value of the stress
energy tensor for a scalar field is given by \cite{Birrell:1982ix}
\begin{equation}\label{traceding}
    \langle 0|T^\mu{}_\mu|0\rangle = \Big(\f{2-D}{4}+(D-1)\xi\Big)\Box
    i\Delta(x;x),
\end{equation}
we see that indeed we have that
\begin{equation}
\f{i}{2}\frac{\delta }{\delta a}{\rm{Tr}}
      \ln\big(\hat{\mathcal{D}}_\chi)= Va^{D-1}g^{\mu\nu}\langle
    T_{\mu\nu}\rangle\,.
\end{equation}
Comparing this with Eq.~(\ref{effectex}) we see that this is
exactly what one expects, justifying thus our procedure. The above
example shows that indeed one can use the functional derivatives
with respect to $a$ to evaluate the quantity that interests us. It
also shows that the rescaling (\ref{chi:rescale}) does not
influence the final result in the context of dimensional
regularization. Notice that (\ref{effectex}) is just an equation
of motion. Therefore, \emph{after} the variation is performed, in
the second line of (\ref{example1}), we can safely evaluate all
the quantities appearing \emph{on-shell}. Hence we only need the
propagators $i\Delta(y;z)$ \emph{on-shell}. This justifies our
on-shell diagonalization procedure, based on which we constructed
the propagators.
\\
Now we apply this technique to the case at hand. The only
difference is that there are more quantum fields. This however
does not change the procedure. Up to an irrelevant constant, the
effective action can be obtained from
\begin{equation}
    \exp[i \Gamma]=\int \big(\mathcal{D}h_{\mu\nu}\big)
              \big(\mathcal{D}\phi\big)\big(\mathcal{D}U^\alpha\big)\big(\mathcal{D}
    \bar{U}^\alpha\big)\exp\Big(i\big(S^{(0)}+S^{(2)}\big)\Big)
\,,\label{Gamma:formal}
\end{equation}
where $U$ and $\bar U$ denote the unrescaled ghost fields
associated with the graviton field $h_{\mu\nu}$. When written in
terms of the rescaled fields this can be recast as ~\footnote{Our
field redefinition~(\ref{fieldredef}) has a Jacobian equal to one.
Furthermore, the rescaling by $a^2$ of $\psi$ with respect to the
'true' graviton will contribute to the effective action as a
$D$-dimensional delta function $\delta^D(0)$. Such a term does not
contribute in dimensional regularization.}
\begin{equation}\label{effective1}
    \begin{split}
    \exp[i \Gamma]&=\int \big(\mathcal{D}z_{ij}\big)\big(\mathcal{D}
    z_{0i}\big)\big(\mathcal{D}
    z_{00}\big)\big(\mathcal{D}\phi\big)\big(\mathcal{D}V^\alpha\big)\big(\mathcal{D}
    \bar{V}^\alpha\big)\exp\Big(i\big(S^{(0)}+S^{(2)}\big)\Big)\\
    &=\exp\Big(i
    S^{(0)}\Big)\f{\mathcal{D}_{\alpha\beta}^{\rm ghost}}{\sqrt{\det(\mathcal{D}^{ij}_{\rm vector})\det( G^{ijkl})}}
\,,
    \end{split}
\end{equation}
where
\begin{equation}
    \begin{split}
        S^{(0)}&=
       \int d^Dx\sqrt{-g}\Big( \f{1}{\kappa}(R-(D-2)\Lambda)
                     -\f{1}{2}(\partial\Phi)^2-V(\Phi)\Big)\\
        S^{(2)}&=\int d^Dx\mathcal{L}^{(2)}\,,
    \end{split}
\end{equation}
and $\mathcal{L}^{(2)}$ is given in (\ref{lagofshell}) and
$(\partial\Phi)^2=-{\Phi'}^2$.
From Eq.~(\ref{effective1}) we obtain
\begin{equation}\label{effective5}
    \begin{split}
    \Gamma&=S^{(0)}+\frac{i}{2}\mathrm{Tr}\ln[\mathcal{D}^{ij}_{\rm vector}]+\frac{i}{2}\mathrm{Tr}\ln[G^{ijkl}]-i\mathrm{Tr}
    \ln[\mathcal{D}^{\rm ghost}_{\alpha\beta}]\\
    &\equiv S^{(0)}+\Gamma_{1L}
\, ,
    \end{split}
\end{equation}
from which we find the expression equivalent to (\ref{effectex})
to be
\begin{equation}\label{effect}
    \begin{split}
        \frac{\delta \Gamma}{\delta a}&=\frac{\delta S_0}{\delta a}+\frac{\delta \Gamma_{1L}}{\delta
        a}\\
        &=Va^{D-1}\bigg[\frac{D(D\!-\!1)(D\!-\!2)}{\kappa}
          \Big(H^2-\frac{1}{D\!-\!1}\Lambda+\frac2{D} a^{-1}H'\Big)
         + (D\!-\!1)p_M-\rho_M\bigg]+
        \frac{\delta \Gamma_{1L}}{\delta a}
        \,,
    \end{split}
\end{equation}
where $p_M$ and $\rho_M$ are the pressure and energy density
associated to the background scalar field matter, given by
$\rho_M=\frac{1}{2}\dot{\Phi}^2+V(\Phi)$ and
$p_M=\frac{1}{2}\dot{\Phi}^2-V(\Phi)$.

 The one loop contribution can be written analogously to
(\ref{example1}) as
\begin{equation}\label{oneloopcontrib}
    \f{\delta \Gamma_{1L}}{\delta a}=\int d^Dx
     \left(\frac{i}{2}[{}_i\Delta_j^{\rm vector}](x;x)
                 \f{\delta}{\delta a}\mathcal{D}^{ij}_{\rm vector}
     + \frac{i}{2}[{}_{ij}\mathcal{M}_{kl}]\f{\delta}{\delta a}G^{ijkl}
     - i[{}^\alpha\Delta_{\rm ghost}^\beta](x;x)
           \f{\delta}{\delta a}\mathcal{D}^{\rm ghost}_{\alpha\beta}
     \right)
\,.
\end{equation}
The functional derivatives should naturally be taken on the
\emph{off-shell} kinetic operators. However, as soon as these
derivatives are taken, we are simply left with an equation of
motion. It is therefore completely valid to impose the background
equations of motion at that point. Therefore the propagators in
(\ref{oneloopcontrib}) can be evaluated \emph{on-shell} and thus
we can use the propagators as they are calculated in the previous
section.\\
Instead of using the kinetic operators as given in
(\ref{kinetic}), we rescale all of our fields as
\begin{equation}\label{rescalegrav}
    z_{\mu\nu}\rightarrow a^{1-D/2}\hat{z}_{\mu\nu}\quad;
             \qquad\phi\rightarrow a^{1-D/2}\hat{\phi}
\,,
\end{equation}
which is identical to the rescaling in Eq.~(\ref{chi:rescale}).
This rescaling changes the effective action (\ref{effective1}) by
a $D$ dimensional coincident delta function that does not
contribute in dimensional regularization. With the following
identity
\begin{equation}
    \phi\sqrt{-g}\Box_s\phi
       = \hat\phi\Big[\eta^{\alpha\beta}\partial_\alpha\partial_\beta
                   + \frac{D-2}{2}\Big(\frac{D}{2}-\epsilon\Big)a^2H^2\Big]\hat\phi
\end{equation}
we can easily calculate the kinetic operators of the rescaled
fields. We will indicate these rescaled kinetic operators with a
hat. The associated propagators are also easily obtained:
\begin{equation}
    \Delta(x;\tilde{x})=
    (a\tilde{a})^{1-D/2}\hat{\Delta}(x;\tilde{x})
\,.
\end{equation}
We now shall consider one functional derivative in detail.  The
others are calculated similarly. We follow the same procedure as
in the example considered at the start of this section. Our
calculation proceeds analogously as in Eq.~(\ref{example1}). We
once again insert the explicit arguments $\eta=x^0$ and
$\tilde{\eta}$ for the two coordinates respectively
\begin{equation}\label{funct1}
    \begin{split}
        \int d^D x&\, {}_{ij}\hat{\Delta}_{kl}(x;x)\f{\delta}{\delta
        a(\eta)}\hat{\mathcal{D}}_{\rm tensor}^{ijkl}(\tilde{\eta})\\
        &=\int  d^D x\, {}_{ij}\hat{\Delta}_{kl}(x;x)\f{\delta}{\delta
        a(\eta)}\Bigg(\eta^{\alpha\beta}\partial_\alpha\partial_\beta
           +\f{D-2}{2}\big(\f{D}{2}-\epsilon(\tilde{\eta})\big)a(\tilde{\eta})^2
        H(\tilde{\eta})^2-a(\tilde{\eta})^2\mathcal{W}\Bigg)
\\
     &\qquad\times\;  \Big(\f{1}{2}\delta^{ik}\delta^{jl}-\f{1}{4}\delta^{ij}\delta^{kl}\Big)
\\
        &=\int  d^D x\, {}_{ij}\hat{\Delta}_{kl}(x;x)\f{\delta}{\delta
        a(\eta)}\Bigg(\eta^{\alpha\beta}\partial_\alpha\partial_\beta
         -\f{D-2}{4}\Big((3D-16)\big(\f{a(\tilde{\eta})'}{a(\tilde{\eta})}\big)^2
           + 6\f{a(\tilde{\eta})''}{a(\tilde{\eta})}\Big)\\
        &\qquad+(D-2)\Lambda
        a(\tilde{\eta})^2-\kappa\Big(\f{1}{2}\Phi'^2-a(\tilde{\eta})^2
        V(\Phi)\Big)\Bigg)\Big(\f{1}{2}\delta^{ik}\delta^{jl}-\f{1}{4}\delta^{ij}\delta^{kl}\Big)
    \end{split}
\end{equation}
\begin{equation}\nonumber
    \begin{split}
        &=\int d^D x\,{}_{ij}\hat{\Delta}_{kl}(x;x)\Bigg(-\f{D-2}{4}\Big(2(3D-16)\big(\f{a'(\tilde{\eta})}
                     {a(\tilde{\eta})^2}\f{\partial}{\partial\tilde{\eta}}
           -\f{a'(\tilde{\eta})^2}{a(\tilde{\eta})^3}\big)
\\
        &\qquad +\;6\big(\f{1}{a(\tilde{\eta})}\f{\partial^2}{(\partial\tilde{\eta})^2}-\f{a(\tilde{\eta})''}{a(\tilde{\eta})^2}\big)\Big)\\
        &\qquad+2(D-2)\Lambda
        a(\tilde{\eta})+2\kappa V(\Phi)
        a(\tilde{\eta})\Bigg)\Big(\f{1}{2}\delta^{ik}\delta^{jl}-\f{1}{4}\delta^{ij}\delta^{kl}\Big)\delta(\tilde{\eta}-\eta)
\\
        &=\int d^D x\,\delta(\tilde{\eta}-\eta)\Bigg(
                -\f{3(D-2)}{2}\f{1}{a(\tilde{\eta})}\f{\partial^2}{(\partial\tilde{\eta})^2}
\\
        &\qquad +\;\f{1}{2}(D-2)(3D-10)\Big(H\f{\partial}{\partial\tilde{\eta}}+(1-\epsilon)
        H(\tilde{\eta})^2a(\tilde{\eta})\Big)\\
        &\qquad+2(D-2)\Lambda
        a(\tilde{\eta})+2\kappa V(\Phi)a(\tilde{\eta})\Bigg)\Big(\f{1}{2}\delta^{ik}\delta^{jl}-\f{1}{4}\delta^{ij}\delta^{kl}\Big){}_{ij}\hat{\Delta}_{kl}(x;x)    \end{split}
\end{equation}
\begin{equation}\nonumber
    \begin{split}
      &  \hskip -0.9cm
 =V\Bigg(-\f{3(D-2)}{2}\f{1}{a}\partial_\eta^2+\f{1}{2}(D-2)(3D-10)H\partial_\eta
\\
&+\,\f{1}{2}(D-2)^2(7-3\epsilon)H^2a\Bigg)
     \Big(\f{1}{2}\delta^{ik}\delta^{jl}-\f{1}{4}\delta^{ij}\delta^{kl}\Big){}_{ij}\hat{\Delta}_{kl}(x;x)
\,.
    \end{split}
\end{equation}
In step three we integrated by parts and dropped the boundary
terms because of the delta function. In step four we used the
background equations of motion (\ref{backgrond1}) for $\Phi$. This
is justified, since corrections to those equations will be of
order one loop and the above expression is already at order one
loop. Therefore the error
one is making is of two loop order.\\
The other functional derivatives we need are calculated similarly
\begin{equation}\label{funct2}
    \begin{split}
        \f{1}{V}\int &{}_i\hat{\Delta}_{j,{\rm{vector}}}(x;x)\f{\delta}{\delta
        a}\hat{\mathcal{D}}_{\rm vector}^{ij}=\Bigg(\f{D-2}{2}\f{1}{a}\partial_\eta^2-\f{1}{2}(D-2)(3D-2)H\partial_\eta\\
        &\qquad-\f{1}{2}(D-2)\Big((3D+2)(1-\epsilon)+4(D-2)\Big)aH^2\Bigg)\delta^{ij}{}_i\hat{\Delta}_{j,{\rm{vector}}}(x;x)\\
    \f{1}{V}\int &\hat{\mathcal{M}}_{(1,1)}(x;x)\f{\delta}{\delta
    a}\hat{\mathcal{D}}_{\rm scalar}=\f{D-2}{2(D-3)}\Bigg(\f{D+2}{2}\f{1}{a}\partial_\eta^2
+\f{3}{2}(D^2-4)H\partial_\eta
\\
&\qquad+\;\f{1}{2}(D-2)\Big((7D+2)-(3D+10)\epsilon\Big)H^2
    a\Bigg)\hat{\mathcal{M}}_{(1,1)}(x;x)\\
    \f{1}{V}\int &\hat{\mathcal{M}}_{(2,2)}(x;x)\f{\delta}{\delta
    a}\hat{\mathcal{D}}_\phi =\Bigg(\f{D-2}{2}\f{1}{a}\partial_\eta^2
       -\f{1}{2}(D-2)^2H\partial_\eta-\Big(\f{1}{2}(D-2)^2(1-\epsilon)
\\
&\qquad+\; 4(D-1-\epsilon)\epsilon\Big)a
H^2\bigg)\hat{\mathcal{M}}_{(2,2)}(x;x)
    \end{split}
\end{equation}
\begin{equation}\nonumber
    \begin{split}
    \f{1}{V}\int &{}^\alpha\hat{\Delta}^\beta_{\rm ghost}\f{\delta}{\delta
    a}\hat{\mathcal{D}}^{\rm ghost}_{\alpha\beta}=\Big(\f{D-2}{2}\f{1}{a}\partial_\eta^2-\f{1}{2}(D-2)^2H\partial_\eta-\f{1}{2}(D-2)^2(1-\epsilon)H^2
    a\Big)\eta_{\alpha\beta} {}^\alpha\hat{\Delta}^\beta_{\rm ghost}\\
    &\qquad+\Big((D-2)\f{1}{a}\partial_\eta^2+4(D-2)H\partial_\eta+4(D-2)(1-\epsilon)H^2
    a\Big)\delta_\alpha^0\eta_{\beta 0} {}^\alpha\hat{\Delta}^\beta_{\rm ghost}\\
    \f{1}{V}\int &\hat{\mathcal{M}}_{(1,2)}(x;x)\f{\delta}{\delta
    a}\sqrt{\kappa}\Big(\f{2}{D-3}\mathcal{Z}-\Phi''\Big)\\
                   &=\f{\sqrt{2(D-2)\epsilon}}{D-3}\Big((D-2)H\partial_\eta+(3D-4-D\epsilon)H^2
    a\Big)\hat{\mathcal{M}}_{(1,2)}(x;x)\,.
    \end{split}
\end{equation}
Here we indicated with $\hat{\mathcal{M}}_{(n,m)}(x;x)$ the
$(n,m)$ component of the propagator matrix $\hat{\mathcal{M}}$
(\ref{props2}).
 The last thing we need before we can calculate the one loop
contribution (\ref{oneloopcontrib}) are the propagators at
coincidence and their derivatives. Since all propagators are
related to the propagator $\hat{\Delta}_n$, we only need that one.
\subsubsection{Infinite volume contribution}\label{sec_infvol}
The propagators that we use are split in two parts
(\ref{final_cutoff_prop}): the infinite volume part and the
corrections due to the infrared cut-off.

We first consider the infinite volume contribution
(\ref{Tomislav}). Taking the $y\rightarrow 0 $ limit of
(\ref{Tomislav}) and dropping the $D$ dependent powers of $y$ that
do not contribute in dimensional regularization we obtain (we drop
the subscript $\infty$)
\begin{equation}\label{propsderiv}
    \begin{split}
    i\hat{\Delta}_n(x;x)=a^{D-2}i\Delta_{n}(x;x)&= |1-\epsilon|^{D-2}
(aH)^{D-2}\f{\Gamma(1-\f{D}{2})}{(4\pi)^{\f{D}{2}}}
\f{\Gamma(\f{D-1}{2}+\nu_{D,\,n})\Gamma(\f{D-1}{2}-\nu_{D,\,n})}
{\Gamma(\f{1}{2}+\nu_{D,\,n})\Gamma(\f{1}{2}-\nu_{D,\,n})}\\
\f{\partial}{\partial\eta}i\hat{\Delta}_{n}(x;x)&=
Ha(D-2)(1-\epsilon)i\hat{\Delta}_{n}(x;x)\\
\Big(\f{\partial}{\partial\eta}\Big)^2i\hat{\Delta}_{n}(x;x)&=H^{2}a^{2}
(D-1)(D-2)(1-\epsilon)^2\,i\hat{\Delta}_{n}(x;x).
    \end{split}
\end{equation}
We can now collect all the terms of (\ref{oneloopcontrib}). Using
(\ref{props1}), (\ref{props2}), (\ref{funct1}), (\ref{funct2}) and
(\ref{propsderiv}) we obtain for the vector and the ghost
\begin{equation}\label{vectghost}
    \begin{split}
    \f{1}{V}\int \frac{i}{2}{}_i\hat{\Delta}^{\rm vector}_j(x;x)\f{\delta}{\delta
    a}\hat{\mathcal{D}}^{ij}_{\rm vector}&=\f{1}{4}(D-1)(D-2)\Big[2(D-1)(D+2)
\\
 &\quad-\,(D^2+D+2)\epsilon-(D-1)(D-2)\epsilon^2\Big]aH^2i\hat{\Delta}_1(x;x)\\
    \f{1}{V}\int -i{}^\alpha\hat{\Delta}_{\rm ghost}^\beta(x;x)\f{\delta}{\delta
    a}\hat{\mathcal{D}}^{\rm ghost}_{\alpha\beta}=&H^2 a\f{1}{2}(D-1)^2(D-2)^2\epsilon(1-\epsilon)i\hat{\Delta}_0(x;x)\\
        &\hskip -1.3cm
  -\f{1}{2}(D-1)(D-2)(1-\epsilon)\Big(2(D+2)-3(D-2)\epsilon\Big)aH^2i\hat{\Delta}_1(x;x)
\,.
    \end{split}
\end{equation}
The last term evaluates to
\begin{equation}\label{thing}
\begin{split}
    &\f{1}{V}\int \f{i}{2}{}_ij\hat{\mathcal{M}}_{kl}(x;x)\f{\delta}{\delta
        a} \hat{G}^{ijkl}=\f{1}{V}\int \f{i}{2}{}_{ij}\tilde{\Delta}_{kl}(x;x)\f{\delta}{\delta
    a}\hat{\mathcal{D}}^{ijkl}_{\rm tensor}\\
    &\hskip 4cm
     +\;\f{1}{2}\lambda^2\Big(\cos(\theta)^2i\hat{\Delta}_0
      +\sin(\theta)^2i\hat{\Delta}_2\Big)\f{\delta}{\delta a}
             \hat{\mathcal{D}}_{\rm scalar}\\
    &\hskip 4cm
      +\;\lambda\cos(\theta)\sin(\theta)(i\hat{\Delta}_0-i\hat{\Delta}_2)
     \f{\delta}{\delta a}\Big(\sqrt{\kappa}\f{2}{D-3}\mathcal{Z}-\sqrt{\kappa}\Phi'')\Big)
\\
    &\hskip 4cm +\;\f{1}{2}
       \Big(\sin(\theta)^2i\hat{\Delta}_0+\cos(\theta)^2i\hat{\Delta}_2\Big)
         \f{\delta}{\delta a}\hat{\mathcal{D}_\phi}
\,.
    \end{split}
\end{equation}
The contribution from the tensor is
\begin{equation}\label{tensor}
    \f{1}{8}D(D-1)(D-2)^2\Big[(1+3D)-3(D-1)\epsilon\Big]
               \epsilon aH^2i\hat{\Delta}_0(x;x)
\,,
\end{equation}
while the terms in (\ref{thing}) multiplying $i\hat{\Delta}_0$
contribute as
\begin{eqnarray}
 &&\f{\epsilon}{4(D-3+\epsilon)}
  \Big[-(D-1)(D-3)(D^2-8D+4)+[D(D-2)(D^2-11D+14)+8]\epsilon
\nonumber\\
 &&\hskip 2.5cm
 +\;(D-1)(D-2)(D+2)\epsilon^2\Big]aH^2i\hat{\Delta}_0(x;x).
\end{eqnarray}
Finally the terms in~(\ref{thing}) multiplying $i\hat{\Delta}_2$
contribute as
\begin{equation}\label{rest}
    \begin{split}
    &\f{1}{4(D-3+\epsilon)}\Big[4(D-1)(D-2)(D-3)(D+3)
                               -\Big(5D^4-20D^3-9D^2+68D-36\Big)\epsilon\\
        &\qquad+
     \Big(D^4-5D^3-4D^2+24D-32\Big)\epsilon^2+\Big(D^3-5D^2+8D+4\Big)\epsilon^3\Big]
                  aH^2i\hat{\Delta}_2(x;x)
\,.
        \end{split}
\end{equation}
Putting (\ref{vectghost}--\ref{rest}) and~(\ref{oneloopcontrib})
together and expanding the result around $D=4$ we obtain the
following non renormalized one loop contributions to the Friedmann
trace equation,

\begin{equation}\label{unrenorm0}
    \begin{split}
        \f{1}{Va^{D-1}H^4}\f{\delta\Gamma^{0}_{1L}}{\delta a}
        &=\f{\epsilon(63\epsilon^2+2\epsilon-105)}{64\pi^2(1+\epsilon)}(1-\epsilon)^2(4\nu_0^2-1)\Bigg(\f{2}{D-4}+\gamma_E+\ln\Big(\f{(1-\epsilon)^2H^2}{4\pi\mu^2}\Big)\\
        &\qquad+\psi(\f{1}{2}+\nu_0)+\psi(\f{1}{2}-\nu_0)+4\f{4\nu_0\nu_0'-1}{4\nu_0^2-1}\Bigg)\\
            &+\f{\epsilon}{32\pi^2(1+\epsilon)^2}(1-\epsilon)^2(4\nu_0^2-1)\Big(93\epsilon^3+90\epsilon^2-169\epsilon-122\Big)\\
        \f{1}{Va^{D-1}H^4}\f{\delta\Gamma^{1}_{1L}}{\delta a}
        &=\f{3(9\epsilon^2-7\epsilon-6)}{64\pi^2}(1-\epsilon)^2(4\nu_1^2-1)\Bigg(\f{2}{D-4}+\gamma_E+\ln\Big(\f{(1-\epsilon)^2H^2}{4\pi\mu^2}\Big)\\
        &\qquad+\psi(\f{1}{2}+\nu_1)+\psi(\f{1}{2}-\nu_1)+4\f{4\nu_1\nu_1'-1}{4\nu_1^2-1}\Bigg)\\
            &+\f{1}{64\pi^2}(1-\epsilon)^2(4\nu_1^2-1)\Big(51\epsilon^2-17\epsilon-54\Big)\\
        \f{1}{Va^{D-1}H^4}\f{\delta\Gamma^{2}_{1L}}{\delta a}
        &=-\f{(5\epsilon-6)(\epsilon^2-2\epsilon-7)}{64\pi^2(1+\epsilon)}(1-\epsilon)^2(4\nu_2^2-1)\Bigg(\f{2}{D-4}+\gamma_E+\ln\Big(\f{(1-\epsilon)^2H^2}{4\pi\mu^2}\Big)\\
        &\qquad+\psi(\f{1}{2}+\nu_2)+\psi(\f{1}{2}-\nu_2)+4\f{4\nu_2\nu_2'-1}{4\nu_2^2-1}\Bigg)\\
            &+\f{1}{64\pi^2(1+\epsilon)^2}(1-\epsilon)^2(4\nu_2^2-1)\Big(3\epsilon^4+13\epsilon^3-83\epsilon^2+35\epsilon+30\Big).
    \end{split}
\end{equation}
Here $\nu_n$ indicates the index $\nu_n$ given in (\ref{scalnun})
with $D=4$ and $\nu'_n$ is $\f{d}{d D}\nu_{D,n}\Big|_{D=4}$.
$\gamma_E$ is the Euler constant and we used the following
expansions of the propagators~(\ref{propsderiv})
\begin{eqnarray}\label{props:expand}
   aH^2 i\hat{\Delta}_n(x;x)&=& a^{D-1}|1-\epsilon|^{D-2}
H^D\f{\Gamma(1-\f{D}{2})}{(4\pi)^{\f{D}{2}}}
\left(\Big(\frac{D-3}{2}\Big)^2-\nu_n^2\right)
\nonumber\\
&\times& \left[1+\frac{D-4}{2}\left(
                           \psi\Big(\frac12+\nu_n\Big)+\psi\Big(\frac12-\nu_n\Big)
                     \right)\right]
\\
|1-\epsilon|^{D-2}H^D\f{\Gamma(1-\f{D}{2})}{(4\pi)^{\f{D}{2}}}
&=&\frac{(1-\epsilon)^2H^4}{16\pi^2}\nonumber\\
&\times&\left(\frac{2\mu^{D-4}}{D-4}+\gamma_E-1+\ln\Big(\frac{H^2}{\mu^2}\Big)
               +\ln\Big(\frac{(1-\epsilon)^2}{4\pi}\Big)
    \right),
\end{eqnarray}
plus terms that vanish in $D=4$. Here
$\psi(z)=(d/dz)\ln(\Gamma(z))$ is the digamma function and $\mu$
is an arbitrary renormalization scale introduced for later
convenience. If we use the explicit expression for $\nu_n$, we can
add all terms together and obtain for the functional derivative of
the effective action
\begin{equation}\label{unrenorm}
    \begin{split}
        \f{1}{Va^{D-1}H^4}\f{\delta\Gamma_{1L}}{\delta a}&=\f{1}{Va^{D-1}H^4}\f{\delta\Gamma^{0}_{1L}}{\delta
        a}+\f{1}{Va^{D-1}H^4}\f{\delta\Gamma^{1}_{1L}}{\delta a}+\f{1}{Va^{D-1}H^4}\f{\delta\Gamma^{2}_{1L}}{\delta
        a}\\
        &=-\f{\epsilon\Big(186-149\epsilon-11\epsilon^2+10\epsilon^3\Big)}{8\pi^2}
                \f{\mu^{D-4}}{D-4}\\
        &-\;\f{\epsilon}{16\pi^2}
       \Bigg[\Big(108+62\epsilon-153\epsilon^2+27\epsilon^3\Big)
            +\Big(186-149\epsilon-11\epsilon^2+10\epsilon^3\Big)\\
        &\qquad\quad\times\,\bigg(
                      \gamma_E+\ln\Big(\f{(1-\epsilon)^2 H^2}{4\pi\mu^2}\Big)
                    \bigg)\Bigg]\\
        &+\f{\epsilon(63\epsilon^2+2\epsilon-105)}{64\pi^2(1+\epsilon)}(1-\epsilon)^2(4\nu_0^2-1)\Bigg(\psi(\f{1}{2}+\nu_0)+\psi(\f{1}{2}-\nu_0)\Bigg)\\
        &+\f{3(9\epsilon^2-7\epsilon-6)}{64\pi^2}(1-\epsilon)^2(4\nu_1^2-1)\Bigg(\psi(\f{1}{2}+\nu_1)+\psi(\f{1}{2}-\nu_1)\Bigg)\\
        &-\f{(5\epsilon-6)(\epsilon^2-2\epsilon-7)}{64\pi^2(1+\epsilon)}(1-\epsilon)^2(4\nu_2^2-1)\Bigg(\psi(\f{1}{2}+\nu_2)+\psi(\f{1}{2}-\nu_2)\Bigg)
\,.\\
    \end{split}
\end{equation}
We kept the digamma functions in terms of $\nu$, since then it
will be more clear how to add the correction terms to regulate the
infrared. In (\ref{unrenorm}), all infrared power-law divergences
are (incorrectly!) subtracted by the automatic subtraction of
dimensional regularization. The infrared logarithmic divergences
are still there and they appear through the poles in the digamma
functions. These issues will be corrected by adding the correction
terms, but before adding the correction terms, we shall first
renormalize the expression (\ref{unrenorm}).

\subsubsection{Renormalization} \label{Renormalization}

The contribution~(\ref{unrenorm}) contains a $1/(D\!-\!4)$
divergence and therefore needs to be renormalized. If we take our
approximation that $\epsilon =$ constant literally, the divergence
is of the form constant $\times H^4 a^{D-1}$. However in a more
realistic treatment, $\epsilon$ is a dynamical parameter and our
result is expected to be correct up to zeroth order in
$\dot{\epsilon}$. Such an approach, using a space-time of locally
constant $\epsilon$ is a generalization of the often considered
locally de Sitter space-time
~\cite{Woodard:2003vp,Woodard:2004ut}. In this more realistic
case, the $\epsilon$ structure of the divergent term should be
taken into account and subtracted accordingly. Therefore if
$\epsilon$ is varying slowly enough, such that (\ref{unrenorm})
remains approximately valid, we still need a counter lagrangian
that produces the same $\epsilon$ structure as the divergence in
(\ref{unrenorm}), in order for the theory to be renormalized at
all times. For this purpose many terms can be used, and indeed
many terms are reported in the literature
~\cite{Christensen:1979iy,HV,Barvinsky:1993zg}. However we cannot
simply use these terms, since counterterms are dependent on the
gauge fixing used ~\cite{Kallosh:1978wt} and as far as we know, no
general calculations have been done using our gauge fixing
(\ref{gauge}). Calculations using (\ref{gauge}) have been done
however in the special case of de Sitter space
~\cite{Tsamis:2005je,Kahya:2007bc}, considering both scalar and
graviton loops. From these works and also e.g. from
~\cite{Finelli:2004bm} it follows that the one loop contribution
in this limit should be finite. Since the de Sitter limit is
$\epsilon\rightarrow 0$, this agrees with (\ref{unrenorm}). To
ensure that the one loop contribution due to gravitons in the de
Sitter case vanishes we need in that limit a counterterm
\begin{equation} \label{dsct1}
     \sqrt{-g} \Big(H^2-\f{\Lambda}{D-1}\Big)^2
\,,
\end{equation}
which in our more general case becomes
\begin{equation}
    \mathcal{L}_{CT1} = \sqrt{-g}\, a_0 \Big(R-\f{D}{D-2}\Big(\kappa
    V(\Phi)+(D-2)\Lambda\Big)\Big)^2
\,,
\end{equation}
where $a_0$ is a constant. This follows from the fact that the
cosmological constant can always be seen as a part of the scalar
potential and thus for an invariant counterterm they should come
together.
 Moreover from ~\cite{Kahya:2007bc} it follows that we also need a counterterm
of the form
   $ \sqrt{-\hat g}H^2\kappa
    \hat g^{\mu\nu}(\partial_\mu\hat\phi)(\partial_\nu\hat\phi)$
to vanish. From Eq.~(\ref{curvature invariants}) it follows that
for our case this generalizes to
\begin{equation}\label{dsct2}
   \mathcal{L}_{CT2} = \sqrt{-g}\,  a_1(R g^{\mu\nu}-D
    R^{\mu\nu})\kappa (\partial_\mu\Phi)(\partial_\nu \Phi)
\,,
\end{equation}
with $a_1$ a constant. Finally, also from ~\cite{Kahya:2007bc} it
follows that a counterterm
 $ \sqrt{-g}\kappa (\Box \Phi)^2$ should not appear.

 Therefore a reasonable choice for the counter-lagrangian is
\begin{equation}\label{counter}
    \begin{split}
    \mathcal{L}_c=\sqrt{-g}\bigg[&a_0\Big(R-\f{D}{D-2}(\kappa
    V+(D-2)\Lambda)\Big)^2+a_1\Big(R g^{\mu\nu}-D
    R^{\mu\nu}\Big)\kappa\partial_\mu\Phi\partial_\nu\Phi\\
    &+a_2\f{\partial^2
    V(\Phi)}{\partial\Phi^2}R+a_3\kappa g^{\mu\nu}(\partial_\mu\Phi)(\partial_\nu\Phi)\f{\partial^2
    V(\Phi)}{\partial\Phi^2}\bigg]
\,.
    \end{split}
\end{equation}
We stress that the counter-lagrangian (\ref{counter}) \emph{for
the purpose of this calculation} could be chosen differently.
There are many other terms with the correct dimensionality that
could have been used~~\cite{Barvinsky:1993zg}. Since the
divergence (\ref{unrenorm}) gives only four constraints (one for
each power of $\epsilon$), we at present can fix only 4
coefficients. This does not mean that the counter-lagrangian is
arbitrary. Different types of calculations could fix the
counter-lagrangian uniquely, as it is for example done in
Ref.~~\cite{HV}. However apart from the two cases mentioned above,
we do not know of any calculation in our gauge, which we could use
to further specify our counter-lagrangian. Thus the 'true'
counter-lagrangian corresponding to the theory will probably
contain different counterterms than (\ref{counter}). However,
these different counterterms do not change the conclusions of this
paper. The only effect would be that in Eq. (\ref{renormalized})
the origin of the $\beta_i$'s changes, but not the fact that they
are essentially arbitrary. The terms in our
counter-lagrangian~(\ref{counter}) contribute as follows to the
Friedmann trace equation
\begin{equation}\label{counterterms}
    \begin{split}
    \f{1}{V}\f{\delta}{\delta a}\int d^Dx &\sqrt{-g} \Big(R-\f{D}{D-2}(\kappa
    V+(D-2)\Lambda)\Big)^2 \\
    &=a^{D-1}H^4(D-2)\epsilon\Big(8\epsilon(2+3\epsilon)+D^2(4+9\epsilon)-2D(2+13\epsilon+12\epsilon^2)\Big)+\mathcal{O}(\epsilon')\\
    &=a^{D-1}H^4\Big(16\epsilon(6+7\epsilon-9\epsilon^2)+4\epsilon(26+37\epsilon-30\epsilon^2)(D-4)+\mathcal{O}((D-4)^2,\epsilon')\Big)
\\
   \f{1}{V}\f{\delta}{\delta a}\int d^Dx&
   \kappa\sqrt{-g}(Rg^{\mu\nu}-DR^{\mu\nu})(\partial_\mu\Phi)(\partial_\nu\Phi)\\
 &=2a^{D-1}H^4(D-1)(D-2)^2\epsilon\Big((D-1)(D-6\epsilon)+6\epsilon^2\Big)
      + \mathcal{O}(\epsilon')\\
 &=a^{D-1}H^4\Big(144\epsilon(1-\epsilon)(2-\epsilon)
         +24\epsilon(23-30\epsilon+8\epsilon^2)(D-4)
            +\mathcal{O}((D-4)^2, \epsilon')\\
    \f{1}{V}\f{\delta}{\delta a}\int d^Dx& \sqrt{-g} \f{\partial^2
    V(\Phi)}{\partial\Phi^2} R\\
 &=a^{D-1}H^4(D\!-\!1)(D\!-\!1\!-\!\epsilon)2\epsilon
  \Big(
          D(D\!-\!2)-2(3D\!-\!4)\epsilon+12\epsilon^2
    \Big) +\mathcal{O}(\epsilon')
\\
     &=a^{D-1}H^4\Big(24\epsilon(3-\epsilon)(2-4\epsilon+3\epsilon^2)
\\
     &\qquad
   +4\epsilon\big(51-88\epsilon+53\epsilon^2-6\epsilon^3\big)(D-4)\Big)
      +\mathcal{O}\Big((D\!-\!4)^2,\epsilon'\Big)
   \end{split}
\end{equation}
\begin{equation}\label{counterterms:2}
    \begin{split}
   \f{1}{V}\f{\delta}{\delta a}\int d^Dx&\kappa\sqrt{-g}
     g^{\mu\nu}(\partial_\mu\Phi)(\partial_\nu\Phi)
     \f{\partial^2 V(\Phi)}{\partial\Phi^2}\\
 &=-4a^{D-1}H^4 (D\!-\!2)^2\epsilon^2(D\!-\!1\!-\!\epsilon)
      +\mathcal{O}(\epsilon')
\\
       &=a^{D-1}H^4\Big(- 16\epsilon^2(3\!-\!\epsilon)   - 16\epsilon^2(4\!-\!\epsilon)(D\!-\!4)\Big)
 +\mathcal{O}\Big((D\!-\!4)^2,\epsilon'\Big)
\,,
   \end{split}
\end{equation}
where we used~(\ref{curvature invariants}) and  once again we used
the background equations of motion~(\ref{backgrond1})
and~(\ref{backgrond2}). We find that all divergencies cancel if
\begin{equation}
    \begin{split}
    a_0=\f{37}{960\pi^2}\f{\mu^{D-4}}{D\!-\!4}+a_0^f\quad
  &;\quad a_1=\f{49}{640\pi^2}\f{\mu^{D-4}}{D\!-\!4}+a_1^f
\\
    a_2=-\f{5}{288\pi^2}\f{\mu^{D-4}}{D\!-\!4}+a_2^f \quad
  &;\quad a_3=-\f{43}{480\pi^2}\f{\mu^{D-4}}{D\!-\!4}+a_3^f
\,,
    \end{split}
\end{equation}
where the $a_i^f$ ($i=0,1,2,3$) indicates a possible finite part.
 Adding the contribution
from the counterterms~(\ref{counterterms}) to the one loop
contribution~(\ref{unrenorm}), we obtain the following
renormalized contribution
\begin{equation}\label{renormalized}
    \begin{split}
        \f{1}{a^3 V}\f{\Gamma_{1L,\rm{ren}}}{\delta a}&=\f{H^4}{16\pi^2}
         \Bigg[\beta_1\epsilon+\beta_2\epsilon^2+\beta_3\epsilon^3
             + \beta_4\epsilon^4\\
&-\epsilon\Big(186-149\epsilon-11\epsilon^2+10\epsilon^3\Big)
       \bigg(\ln\Big(\f{(1-\epsilon)^2H^2}{4\pi\mu^2}\Big)
           \bigg)\\
       &+\f{\epsilon(63\epsilon^2+2\epsilon-105)}{4(1+\epsilon)}(1-\epsilon)^2(4\nu_0^2-1)\bigg(\psi(\f{1}{2}+\nu_0)+\psi(\f{1}{2}-\nu_0)\bigg)\\
        &+\f{3(9\epsilon^2-7\epsilon-6)}{4}(1-\epsilon)^2(4\nu_1^2-1)\bigg(\psi(\f{1}{2}+\nu_1)+\psi(\f{1}{2}-\nu_1)\bigg)\\
        &-\f{(5\epsilon-6)(\epsilon^2-2\epsilon-7)}{4(1+\epsilon)}(1-\epsilon)^2(4\nu_2^2-1)\bigg(\psi(\f{1}{2}+\nu_2)+\psi(\f{1}{2}-\nu_2)\bigg)\Bigg]
           \,.
    \end{split}
\end{equation}
Here the parameters $\beta_i$ ($i=1,2,3,4$) are given in terms of
the finite coefficients $\alpha_i^f$,
\begin{equation}\label{betas}
    \begin{split}
 \beta_1 &= 16\pi^2\times 12\Big[8 a_0^f
                               +24a_1^f
                               +12 a_2^f
                               \Big]
          -186\gamma_E
          + \frac{1727}{3}\\
 \beta_2 &= 16\pi^2\times 4\Big[28a_0^f-108a_1^f
          -84a_2^f-12a_3^f\Big]
          +149\gamma_E
          - \frac{5969}{9}\\
 \beta_3 &= 16\pi^2\times 4\Big[-36a_0^f+36a_1^f
          +78a_2^f+4a_3^f\Big]
          +11\gamma_E
          + \frac{10457}{45}\\
 \beta_4 &= 16\pi^2\Big[-72 a_2^f\Big]
         - 10\gamma_E
         -\frac{61}{3}
\,.
\end{split}
\end{equation}
All $\beta_i$'s ($i=1,2,3,4$) in Eq.~(\ref{renormalized}) are free
parameters that remain undetermined until they are fixed by
experiment.

\subsubsection{Correction terms}
The effective action (\ref{renormalized}) is divergent for half
integer values of the parameters $\nu_i\geq 3/2$ (notice that the
pole at $\nu_i=1/2$ is cancelled due to the pre-factor)
\begin{equation}
    \nu_i = 3/2+N\qquad;\qquad N=\{0,1,2\ldots\}.
\end{equation}
The reason is that the propagators used do not describe the
infrared physics correctly and we need to add the correction terms
$\delta i\Delta_N$ and $\delta i\Delta^N$ as given in
(\ref{corrections2}). For concreteness we shall consider the late
time behavior in an accelerating space-time, since in that case we
expect the most significant backreaction. In an accelerating
space-time, $\eta$ goes to zero at late times, and thus we do not
care about the $\delta i \Delta^N$ corrections, since they quickly
become insignificant in that case. Since the correction terms are
ultraviolet finite, we can put $D=4$ in all terms. Following the
same procedure as in section \ref{sec_infvol}, we find for the
coincident limit and its derivatives
\begin{equation}\label{correct_coinc}
    \begin{split}
    \delta i \hat{\Delta}_{N,n}(x;x) &= A_{N,n} (aH)^2 z_0^{2N+3-2\nu_n}\\
    \f{\partial}{\partial \eta}\delta i
    \hat{\Delta}_{N,n}(x;x)&=-(1+2N-2\nu_n)(1-\epsilon)(aH)\delta i
    \hat{\Delta}_{N,n}(x;x)\\
    \Big(\f{\partial}{\partial \eta}\Big)^2 \delta i
    \hat{\Delta}_{N,n}(x;x)&=(1+2N-2\nu_n)(2N-2\nu_n)(1-\epsilon)(aH)\delta i
    \hat{\Delta}_{N,n}(x;x)\,,\\
\end{split}
\end{equation}
where $\nu_n$ is given in (\ref{scalnun}) and a subscript $n$
implies that the quantity is evaluated with $\nu=\nu_n$. We also
defined
\begin{equation}
A_{N,n}\equiv
-\f{1}{4\pi^{5/2}}\f{1}{3+2N-2\nu_n}\f{\Gamma(\nu_n-N)\Gamma(2\nu_n-N)}{\Gamma(\f{1}{2}+\nu_n
-N)\Gamma(N+1)}(1-\epsilon)^2.
\end{equation}

We find for the corrections due to the vector and the ghost
(\ref{funct2})
\begin{equation}\label{vectghost_cor}
    \begin{split}
    \f{1}{V}\int &\frac{i}{2}(\delta {}_i\hat{\Delta}^{\rm vector}_j)_N(x;x)\f{\delta}{\delta
    a}\hat{\mathcal{D}}^{ij}_{\rm vector}=   3\Bigg(4+7(1-\epsilon)\\
    &-5(1-\epsilon)(1+2N-2\nu_1)-(1-\epsilon)^2(N-\nu_1)(1+2N-2\nu_1)\Bigg)a H^2 \delta i\hat{\Delta}_{N,1}(x;x)\\
    \f{1}{V}\int& -i({}^\alpha\hat{\Delta}_{\rm
    ghost}^\beta)_N(x;x)\f{\delta}{\delta
    a}\hat{\mathcal{D}}^{\rm ghost}_{\alpha\beta}=6(1-\epsilon)(N-\nu_0)\\
    &\times\Big((2\nu_0-2N)(1-\epsilon)+\epsilon-3\Big)a H^2\delta
    i\hat{\Delta}_{N,0}(x;x)\\
    &\quad+6(1-\epsilon)(N-\nu_1)\Big((2\nu_1-2N)(1-\epsilon)+\epsilon+1\Big)aH^2\delta
    i\hat{\Delta}_{N,1}(x;x).
    \end{split}
\end{equation}
The tensor contribution to (\ref{thing}) is
\begin{equation}\label{tensor_cor}
    -6\big(3(\nu_0-N)(1-\epsilon)+2\big)\big(2(\nu_0-N)(1-\epsilon)-3\big) aH^2\delta i\hat{\Delta}_{N,0}(x;x)
\,,
\end{equation}
while the terms in (\ref{thing}) multiplying $\delta
i\hat{\Delta}_{N,0}$ contribute as
\begin{equation}
\begin{split}
    \f{1}{1+\epsilon}\Bigg(&-4(3-\epsilon)\epsilon+(1-\epsilon)^2(1+3\epsilon)(2N^2+2\nu_0^2)-(3+\epsilon)(1-\epsilon)(1-3\epsilon)\\
    &+N(1-\epsilon)\Big(3-4\nu_0-8\epsilon(1+\nu_0)-3\epsilon^2(1-4\nu_0)\Big)\Bigg)aH^2\delta i\hat{\Delta}_{N,0}(x;x).
\end{split}
\end{equation}
Finally the terms in~(\ref{thing}) multiplying $\delta
i\hat{\Delta}_{N,2}$ contribute as
\begin{equation}
\begin{split}
    \f{1}{1+\epsilon}\Bigg(&(1-\epsilon)^2(3+\epsilon)(2N^2+2\nu_1^2)+2(3+5\epsilon(1-\epsilon)+\epsilon^3)\\
    &+(1-\epsilon)(3+\epsilon)\Big((5+\epsilon)(1-N)-4(1-\epsilon)N\nu_1\Big)
    \Bigg)aH^2\delta i\hat{\Delta}_{N,2}(x;x).
\end{split}
\end{equation}
We add the corrections together to obtain the following three
contributions

\begin{equation}\label{unrenorm_corr}
    \begin{split}
        \f{1}{Va^{D-1}H^4}\f{\delta\Gamma^{(0)}_{N}}{\delta
        a}&=\Bigg(-\f{\epsilon(63\epsilon^2+2\epsilon-105)}{1+\epsilon}+\f{1-\epsilon}{1+\epsilon}(3+2N-2\nu_0)\\
        &\qquad\times\Big((\nu_0-N)(1-\epsilon)(23+21\epsilon)+(4+3\epsilon)(3-7\epsilon)\Big)\Bigg)\f{\delta
        i \Delta_{N,0}}{H^2}\\
        \f{1}{Va^{D-1}H^4}\f{\delta\Gamma^{(1)}_{N}}{\delta
        a}&=\Bigg(-3(9\epsilon^2+7\epsilon+6)\\
        &\qquad+9(1-\epsilon)(3+2N-2\nu_1)\Big((\nu_1-N)(1-\epsilon)-\epsilon\Big)\Bigg)\f{\delta
        i \Delta_{N,1}}{H^2}\\
        \f{1}{Va^{D-1}H^4}\f{\delta\Gamma^{(2)}_{N}}{\delta
        a}&=\Bigg(\f{(5\epsilon-6)(\epsilon^2-2\epsilon-7)}{1+\epsilon}\\
        &\qquad-\f{(1-\epsilon)(3+\epsilon)}{1+\epsilon}(3+2N-2\nu_2)\Big((\nu_2-N)(1-\epsilon)+4-\epsilon\Big)\Bigg)\f{\delta
       i \Delta_{N,2}}{H^2},
    \end{split}
\end{equation}
where we have written $\delta \Gamma_N^{(n)}$ for the contribution
to $\delta\Gamma_N$ that multiplies $\Delta_{N,n}$.

From these three contributions we see explicitly that indeed when
$\nu=N+3/2$, the corrections have the correct prefactor to add up
correctly to cancel the divergence in the digamma functions
(\ref{renormalized}). Inserting (\ref{renormalized}) and
(\ref{unrenorm_corr}) into (\ref{effect}) at $D=4$ we obtain the
final one loop corrected Friedmann trace equation
\begin{equation}\label{renormalized2}
    \begin{split}
        &\f{24}{\kappa}\Big((1-\f{1}{2}\epsilon)H^2-\f{1}{3}\Lambda\Big)+3
        p_M-\rho_M+\f{H^4}{16\pi^2}\Bigg\{
        \beta_1\epsilon+\beta_2\epsilon^2+\beta_3\epsilon^3
             + \beta_4\epsilon^4\\
&-\epsilon\Big(186-149\epsilon-11\epsilon^2+10\epsilon^3\Big)
       \ln\Big(\f{(1-\epsilon)^2H^2}{4\pi\mu^2}\Big)
           \\
       &+\f{\epsilon(63\epsilon^2+2\epsilon-105)}{4(1+\epsilon)}(1-\epsilon)^2\\
       &\qquad\times\bigg((4\nu_0^2-1)\Big(\psi(\f{1}{2}+\nu_0)+\psi(\f{1}{2}-\nu_0)\Big)-\sum_N\f{64\pi^2}{(H(1-\epsilon))^2}\delta i\Delta_{N,0}\bigg)\\
        &+\f{3(9\epsilon^2-7\epsilon-6)}{4}(1-\epsilon)^2\\
        &\qquad\times\bigg((4\nu_1^2-1)\Big(\psi(\f{1}{2}+\nu_1)+\psi(\f{1}{2}-\nu_1)\Big)-\sum_N\f{64\pi^2}{(H(1-\epsilon))^2}\delta i\Delta_{N,1}\bigg)\\
        &-\f{(5\epsilon-6)(\epsilon^2-2\epsilon-7)}{4(1+\epsilon)}(1-\epsilon)^2\\
        &\qquad\times\bigg((4\nu_2^2-1)\Big(\psi(\f{1}{2}+\nu_2)+\psi(\f{1}{2}-\nu_2)\Big)-\sum_N\f{64\pi^2}{(H(1-\epsilon))^2}\delta i\Delta_{N,2}\bigg)\Bigg\}\\
        &+\sum_N(3+2N-2\nu_0)\Bigg(\f{1-\epsilon}{1+\epsilon}\Big((\nu_0-N)(1-\epsilon)(23+21\epsilon)+(4+3\epsilon)(3-7\epsilon)\Big)\Bigg)H^2\delta
        i \Delta_{N,0}\\
        &+\sum_N(3+2N-2\nu_1)\Bigg(9(1-\epsilon)\Big((\nu_1-N)(1-\epsilon)-\epsilon\Big)\Bigg)H^2\delta
        i \Delta_{N,1}\\
        &-\sum_N(3+2N-2\nu_2)\Bigg(\f{(1-\epsilon)(3+\epsilon)}{1+\epsilon}\Big((\nu_2-N)(1-\epsilon)+4-\epsilon\Big)\Bigg)H^2\delta
       i \Delta_{N,2}=0
           \,.
    \end{split}
\end{equation}
This equation is our final result of this section. It presents the
one loop quantum corrected Friedmann trace equation, in the
presence of both graviton and scalar fluctuations.

\subsection{Discussion}\label{s_disc}
Having found the one loop corrected Friedmann trace equation, we
can ask the question whether the quantum effects calculated will
have any significant effect. We immediately see from
(\ref{renormalized2}) that the quantum contribution is suppressed
by a factor $H^2/m_p^2$. Because of this suppression, the one loop
contribution can only become relevant for the dynamics of the
Universe if there is a significant enhancement. In other words:
for any significant effect the one loop correction terms should
grow in time with respect to the classical tree level terms. In an
accelerating universe $z_0$ goes to zero at late times. From
(\ref{correct_coinc}) we see that the corrections $\delta i
\Delta_{N,n}$ are proportional to $H^4 z_0^{2N+3-2\nu_n}$ and thus
this contribution might grow in time with respect to the
background (which scales as $H^2$), making the initially tiny
quantum effects significant at late times. The question whether
this growth is significant is equivalent to the question whether
the quantity
\begin{equation}
    H^2 z_0^{3+2N-2\nu}
\end{equation}
grows in time. In an accelerating universe the fastest growing
term has $N=0$.  Since $z_0=-k\eta$ this term grows with $\eta$ as
\begin{equation}
    \propto \eta^{\f{3-\epsilon}{1-\epsilon}-2\nu}.
\end{equation}
Thus we see that for the three different $\nu_n$'s contributing to
(\ref{renormalized2}) this becomes
\begin{equation}
    \begin{split}
    \propto\eta^{\f{3-\epsilon}{1-\epsilon}-2\nu_0}&=\eta^0\\
    \propto\eta^{\f{3-\epsilon}{1-\epsilon}-2\nu_1}&=\eta^2\\
    \propto\eta^{\f{3-\epsilon}{1-\epsilon}-2\nu_2}&=\eta^4.
    \end{split}
\end{equation}
These general expressions fail near the $N=0$ pole of the digamma
functions (meaning $\nu_i = 3/2$). Here one will typically pick up
an additional logarithm, $\ln(z_0)$ from the correction terms.
Thus we see that the $\nu_0$ contribution, in the presence of such
an additional logarithm, actually grows in time. However
$\nu_0=3/2$ implies $\epsilon=0$ and we see from
(\ref{renormalized2}) that this contribution is cancelled by the
pre-factor. Thus none of the contributions actually grows in
time.\\
 The fact that the pre-factor is such that near de Sitter
space the logarithmic growth drops out, is also in concordance
with results obtained in~\cite{Abramo:1998hj}. This property
however does not have to stay true in general. For example if
higher order loop corrections are taken into account, this
structure might change such that while the correction terms are
extremely small, their contribution grows in time, making them
significant at late enough times. Also when considering a
different background geometry, one might get growing effects even
at one loop order. In ~\cite{Abramo:1996gd,Abramo:1997hu} for
example one loop contributions have been considered in chaotic
inflationary models. Although those works found a secular growth,
in ~\cite{Abramo:2001dc} it was concluded that this growth
disappears if one considers truly gauge invariant quantities.

\section{Conclusion}\label{s_con}
In this paper we have calculated the one loop contribution to the
effective Friedmann equations due to graviton and matter quantum
fluctuations. The background space-time we use is FLRW with the
additional constraint that $\epsilon$ is constant. In such a
calculation many delicate issues have to be taken into account.
First of all there is mixing between the gravitational and matter
degrees of freedom. This mixing can (at least on-shell) be removed
by field redefinitions (\ref{fieldredef}) and (\ref{Rotation}).
Secondly, there is the issue of gauge freedom, which we can deal
with in the standard way by adding a gauge fixing term
(\ref{gauge}) and the associated Fadeev-Popov ghost lagrangian
(\ref{ghost_lag}). Finally, to calculate the one loop
contribution, one needs to know the propagators. The propagators
needed will however typically be infrared divergent. To regulate
this infrared divergence we assumed that the spatial sections of
the universe are compact. This effectively implies that the
propagators should be calculated using an infrared cut-off $k_0$.
Using this approach the scalar propagator was constructed in
\cite{Janssen:2008px}. In the present work we have constructed
the graviton and matter propagators by making use of the same method.
All these technical issues can thus be resolved, leading to our
final answer for the one loop Friedmann trace equation (\ref{renormalized2}).
Although the final answer might look intimidating,
we can essentially write it as two types
of contributions. One is  $H^4$ times a constant and the other is
$H^4$ times some power of $z_0=|\eta| k_0$. Since the tree level
contribution is of the order of $H^2 m_p^2$, we see that -- as
expected -- quantum corrections to the Friedmann equation are
suppressed as $H^2/m_p^2$. There are thus only two possibilities
for the quantum corrections to have a significant effect. Either
the coefficient in front of $H^4$ is large, or any of the powers
of $z_0$ grows fast enough to compensate the suppression after a
sufficient amount of time. Looking at equation
(\ref{renormalized2}) we see that for typical values of the
parameters, the coefficient will be of order one. One might think
that the digamma functions might grow large. Indeed, these
functions have a simple pole when the argument is a negative
integer. However, as is shown in ~\cite{Janssen:2008px}, these
poles are due to a logarithmic infrared divergence. These poles
should therefore be subtracted by the correction terms, and in
~\cite{Janssen:2008px} it is shown that this indeed is what
happens. Thus the only possibility for a significant backreaction
of quantum fluctuations on the background space-time is if the
correction terms grow fast enough in time. In section \ref{s_disc}
we showed however that this will not happen. The only possibility
where the powers of $z_0$ grow fast enough is the special case of
de Sitter ($\epsilon=0$). However in this case, the growing power
of $z_0$ drops out due to the vanishing of its prefactor.\\

It is however not clear whether these conclusions will also hold
in more general cases. Generalizations could include for example
studying a more general class of backgrounds, using different
regularization schemes for the infrared. For example
in~\cite{JanssenProk} a scheme using mode matching is introduced.
This scheme gives similar results for accelerating space-times as
the scheme used here, but the results differ for decelerating
space-times. Another generalization would be to include two or
higher loop contributions into account. Indeed studies in de
Sitter space show, for example, that at two loop order the
backreaction due to gravitons might indeed grow in
time~\cite{Tsamis:1996qk,Tsamis:1996qm,Tsamis:1996qq}. This paper
should be seen as a first step in trying to understand
backreaction on a more general background space-time.

\section*{Acknowledgements}

We would like to thank Richard P. Woodard and G. Stavenga for
useful discussions. The authors acknowledge financial support by
Utrecht University. T.P. acknowledges financial support by FOM
grant 07PR2522.

\end{document}